\documentclass[]{raa}            
\usepackage{graphicx,times}
\usepackage{natbib}
\usepackage{amssymb}
\usepackage{ulem}
\usepackage{color}

\providecommand{\bm}[1]{\mbox{\boldmath$#1$\unboldmath}}
\def\kms{$\rm{km~s}^{-1}$}

\begin{document}

   \title{Multicolor Photometry of the Nearby Galaxy Cluster A119}

 \volnopage{ {\bf 2011} Vol.\ {\bf X} No. {\bf XX}, 000--000}
   \setcounter{page}{1}

   \author{Jin-Tao Tian
      \inst{1,2,3}
   \and Qi-Rong Yuan
      \inst{4}
   \and Xu Zhou
      \inst{1,3}
   \and Zhao-Ji Jiang
      \inst{1,3}
   \and Jun Ma
      \inst{1,3}
   \and Jiang-Hua Wu
         \inst{1,3}
   \and Zhen-Yu Wu
      \inst{1,3}
   \and Zhou Fan
      \inst{1,3}
   \and Tian-Meng Zhang
      \inst{1,3}
    \and Hu Zou
      \inst{1,3}
   }

   \institute{National Astronomical Observatories, Chinese Academy of Sciences,
             Beijing 100012, China; {\it tianjintao2010@gmail.com}\\
        \and
             Graduate University, Chinese Academy of Sciences, Beijing 100039, China
        \and
             Key Laboratory of Optical Astronomy, National Astronomical Observatories,
             Chinese Academy of Sciences, Beijing, China
        \and
             Department of Physics, Nanjing Normal University, Nanjing, China
\vs \no
}

\abstract{This paper presents multicolor optical photometry of the
nearby galaxy cluster Abell~119 ($z=0.0442$) with the
Beijing-Arizona-Taiwan-Connecticut (BATC) system of 15 intermediate
bands. Within the BATC viewing field of 58\arcmin $\times$ 58\arcmin,
there are 368 galaxies with known spectroscopic redshifts, including
238 member galaxies (called sample I). Based on the spectral energy
distributions (SEDs) of 1376 galaxies brighter than $i_{\rm
BATC}=19.5$, photometric redshift technique and the color-magnitude
relation of early-type galaxies are applied to select faint member
galaxies. As a result, 117 faint galaxies were selected as new member
galaxies. Combined with sample I, an enlarged sample (called sample
II) of 355 member galaxies is obtained. Spatial distribution and
localized velocity structure for two samples demonstrate that A119 is
a dynamically complex cluster with at least three prominent
substructures in the central region within 1 Mpc. A large velocity
dispersion for the central clump indicates a merging along the line
of sight. No significant evidences for morphology and luminosity
segregations are found in both samples. With the evolutionary synthesis
model PEGASE, environmental effect on the star formation
properties is confirmed. Faint galaxies in low-density region tend to
have longer time scales of star formation, smaller mean stellar ages,
and lower metallicities of interstellar medium, which is in agreement
with the context of hierarchical cosmological scenario.
 }

   \authorrunning{J.-T. Tian et al. }    
   \titlerunning{ Multicolor Photometry of the Galaxy Cluster A119 }  
   \maketitle
\section{Introduction}           
\label{sect:intro} As the most massive gravitationally bound
structures in the universe, clusters of galaxies are not only the
powerful probes tracing the large scale structure, but also the
unique astrophysical laboratories for investigating evolution of
galaxies and dark matter in dense environment
(\citealt{bahcall88,pearce00,cortese04,mantz08}). The hierarchical
formation scenario predicts that galaxy clusters are formed by
continuous accreting subunits of smaller mass at the nodes of
large-scale filaments (\citealt{west91,katz93}). Substructures in
galaxy clusters have been intensively studied since the ROSAT era
using the X-ray surface brightness distribution from observations and
simulations (\citealt{jones84,ventimiglia08,piffaretti08,zhang09}). Based
on quantitative measure of substructure in nearby clusters, a
significant fraction ($>$ 50\%) of clusters shows evidence of merging
(\citealt{geller82,dressler88,mohr95,buote96}), indicating that galaxy
clusters are still dynamically young units, undergoing the process of
formation (\citealt{dressler88,ohara06}).

The Beijing-Arizona-Taiwan-Connecticut (BATC) galaxy cluster survey
aims at obtaining the spectral energy distributions (SEDs) of
galaxies in more than 30 nearby clusters at different dynamic
statuses, with 15 intermediate-band filters in optical band
(\citealt{yuan01}). The photometric redshift technique makes it
possible to pick up faint cluster galaxies down to $V \sim 20$ mag.
With the enlarged sample of member galaxies, dynamical substructures,
luminosity function, luminosity segregation, and star formation
properties of these nearby clusters can be addressed
(\citealt{yuan03,yang04,zhang10,liu11,pan11}). As one of the
brightest extended X-ray sources with multiple clumps, the galaxy
cluster Abell~119 (hereafter A119) is included in the target list of
the BATC cluster survey.

The nearby ($z=0.0442$) cluster A119, located at
00$^{h}$56$^{m}$21$^{s}$.4, -01\degr15\arcmin47\arcsec.0 (J2000.0),
is of richness class 1 (\citealt{abell58}), and is classified as
Bautz-Morgan class II - III (\citealt{bautz70}). The most luminous
member galaxy, UGC~579, is classified as a cD galaxy
(\citealt{postman95, saglia97}). \citet{peres98} argued that there is
no cooling flow in this cluster. What A119 appeals to people are its
high X-ray luminosity and multiple substructures. \citet{edge90}
derived a luminosity of $L_{\rm 2-10keV}\sim 2.58\times10^{44}$ erg
s$^{-1}$ for A119. \citet{fabricant93} argued that substructures in
its core involves at least three clumps from inspection of X-ray maps
and galaxy isopleths. However, \citet{girardi97} argued that A119 is
a regularly shaped cluster, and no evidence of substructure is found
on the basis of their own detection method. There are three discrete
radio galaxies in A119, namely 0053-015, 0053-016, and 3C~29
(\citealt{feretti99}). There is no evidence for luminosity
segregation in A119 (\citealt{pracy05}). Fig.~1 shows the ROSAT and
NVSS smoothed contours superimposed on the $f_{\rm BATC}$-band image
of 58\arcmin $\times$ 58\arcmin centered on the cD galaxy UGC~579.
The X-ray brightness contour shows a clear NE elongation in the
central region, suggesting that A119 is not likely to be a
well-relaxed regular cluster.

The structure of this paper is as follows. In Section 2 we describe
the BATC multicolor photometric observations and data reduction. In
Section 3, spatial distribution and dynamics of cluster galaxies with
known spectroscopic redshifts are investigated. Photometric redshift
technique and its application on selection of faint member galaxies
are given in section 4. In section 5, based on enlarged sample of
cluster galaxies, we unveil some observational properties of A119,
such as spatial distribution, dynamics, morphology-density relation,
and luminosity segregation. In section 6, star formation properties
for the spectroscopically-confirmed member galaxies is presented.
Finally, we summarize our work in Section 7. Throughout this paper,
the $\Lambda$CDM cosmology model with H$_{0}$=73 km s$^{-1}$
Mpc$^{-1}$, $\Omega_{m} $=0.3, and $\Omega_{\Lambda}$=0.7 are
assumed.

\label{sect:Obs}

\begin{figure}[!hbp]
\centering
\includegraphics[width=90mm,height=80mm]{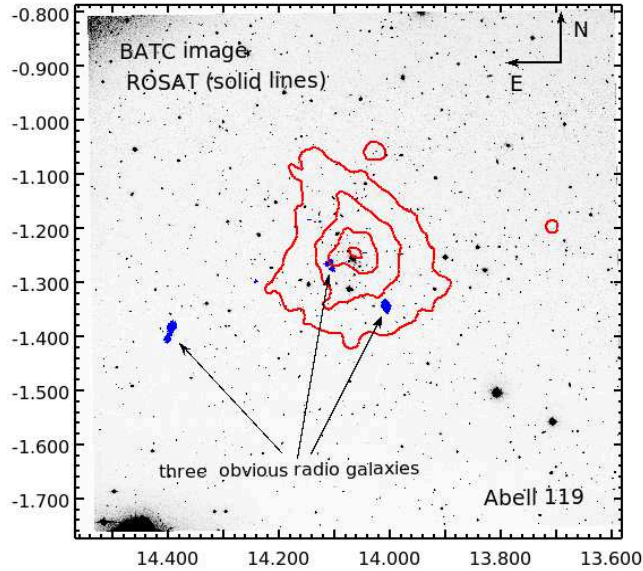}
\caption{The smoothed contours of the ROSAT PSPC image in soft X-ray
band (0.1-2.4keV) ({\sl solid lines}) and the NVSS FIRST (Faint
Images of the Radio Sky at Twenty-cm) map at wavelength of $20 cm$ ,
overlaying on optical image of 58\arcmin $\times$ 58\arcmin in the
BATC $f$ band.The sizes of gaussian smoothing windows are adopted as
30\arcsec and 4\arcmin for the radio and X-ray contours,
respectively. \label{fig1}}
\end{figure}

\section{Observations and Data Reduction}
\label{sect:Obs}

The observation of A119 were carried out with the 60/90cm f/3 Schmidt
Telescope of National Astronomical Observatories of China (NAOC), at
the Xinglong station with an altitude of 900m. The BATC multicolor
photometric system includes 15 intermediate-band filters, namely,
$a-k$, and $m-p$, covering the whole optical wavelength range from
about 3000 to 10000 \AA. These filters are specially designed to
avoid most of bright night-sky emission lines (\citealt{yan00}). The
transmission curves can be seen in \citet{yuan03}.

An 2048 $\times$ 2048 Ford CCD camera was equipped with the BATC
system before October 2006. The field of view was
58\arcmin$\times$58\arcmin, with a scale of 1.7\arcsec per pixel. For
pursuing better spatial resolution and higher sensitivity in blue
bands, a new E2V 4096 $\times$ 4096 CCD is now put into service. This
CCD has a quantum efficiency of 92.2\% at 4000 \AA\, and the field of
view is extended to 92\arcmin$\times$92\arcmin. The spatial scale
becomes 1.36\arcsec per pixel since the pixel size is 12 $\mu$m,
exactly 4/5 of former pixel size (15 $\mu$m). The details of the
telescope, camera, and data-acquisition system can be found elsewhere
(\citealt{zhou01, yan00}).

From September 2002 to November 2006, only 12 BATC filters, from {\it
d} to {\it p}, were taken to target A119 with the old CCD,
discontinuously. Since 2007, the exposures in {\it a, b, c} bands
have been completed with new CCD camera. In total, we have about 54
hr of exposure (see more details in Table 1).
\begin{table}
\bc
\begin{minipage}[]{120mm}
\caption[]{  Parameters of the BATC filters and the observational statistics
of A119 }
\end{minipage}

\small
 \begin{tabular}{ccccccccccc}
  \hline\hline\noalign{\smallskip}
No. & Filter & $\lambda_{eff}^{a}$ &  FWHM  & Exposure &  Number of  & $ Seeing^{\rm b} $   & Completeness  & Objects  \\
   & Name & (\AA) & (\AA) &(second)  & Images &
  (arcsec)  & Magnitude & Detected \\
  \hline\noalign{\smallskip}
 1 & a &  3371 & 356  &  20760  & 19  &  4.18  &  21.0 & 7526  \\
 2 & b &  3894 & 294  &   4860  & 12  &  5.94  &  20.5 & 7471  \\
 3 & c &  4201 & 297  &   8880  & 11  &  4.36  &  20.5 & 7959  \\
 4 & d &  4546 & 367  &  24000  & 23  &  4.15  &  20.5 & 6079  \\
 5 & e &  4872 & 377  &  19800  & 19  &  4.95  &  20.5 & 7745  \\
 6 & f &  5247 & 338  &  12000  & 13  &  4.50  &  20.5 & 6813  \\
 7 & g &  5784 & 285  &   8400  & 10  &  3.51  &  20.0 & 7400  \\
 8 & h &  6073 & 310  &   8100  &  9  &  3.48  &  19.5 & 7730  \\
 9 & i &  6709 & 518  &   3960  &  5  &  3.65  &  19.5 & 8158  \\
10 & j &  7010 & 168  &   6900  &  8  &  4.11  &  19.0 & 7254  \\
11 & k &  7526 & 200  &   8400  & 10  &  3.78  &  19.0 & 6711  \\
12 & m &  8024 & 256  &  15900  & 17  &  3.98  &  19.0 & 7250  \\
13 & n &  8517 & 160  &  12000  & 13  &  4.77  &  19.0 & 7879  \\
14 & o &  9173 & 255  &  25200  & 24  &  4.15  &  18.5 & 8219  \\
15 & p &  9723 & 280  &  22079  & 22  &  4.25  &  18.5 & 7158  \\
  \noalign{\smallskip}\hline
\end{tabular}
\ec
\tablecomments{0.86\textwidth} {$^{a}$ Effective wavelengths of the
filters; $^{b}$ This column lists the seeing of the combined images.}
\end{table}



The standard procedures of bias substraction, dome flat-field
correction, and position calibration were carried out with automatic
data-processing software, PIPLINE I, developed specially for the BATC
multicolor photometry (\citealt{fan96, zhou01}). The cosmic rays and
bad pixels were corrected by comparing the multiple images during
combination.

For detecting sources and measuring the fluxes within a given
aperture in the combined BATC images, we use the photometry package,
PIPELINE II, developed on the basis of DAOPHOT (\citealt{stetson87})
kernel (\citealt{zhou03a}). The objects with signal-to-noise ratio
greater than the threshold 3.5 $\sigma$ in $i$, $j$, and $k$ bands
are considered to be detected. Because the pixel size ratio between
the old and the new CCD is 5:4, an aperture radius of 4 pixels
(i.e.$r$=1.\arcsec7$\times$4=6.\arcsec8) is taken for the images in
12 redder bands ($d$ to $p$), and a radius of 5 pixels (i.e.,
$r$=1\arcsec.36$\times$5=6.\arcsec8) is adopted for the images in
bluer bands ({\it a,b,c}). Flux calibration in the BATC system is
performed by using four Oke-Gunn standard stars (HD 19445, HD 84937,
BD+262606, and BD+17 4708) (\citealt{gunn83}). The procedures of BATC
flux calibration are slightly corrected by (\citealt{zhou01}). Model
calibration on the basis of the stellar SED library are performed to
check the results of flux calibration via standard stars
(\citealt{zhou99}). The flux measurements derived by above these two
calibration methods are in accordance with each other for most
filters. As a result, the SEDs of 10,605 sources have been obtained.

Spatial scale at cluster redshift z=0.0442 is 0.834 kpc/arcsec,
the typical seeing of combined images in the BATC bands is about 4.\arcsec2,
corresponding to 3.5 kpc, which is smaller than the size of a typical spiral
galaxy. For checking the completeness of the BATC detection of galaxies,
we compare the SDSS galaxies down to $r < 19.5$ within a central region with
a radius of 0.5 degree. There are 1121 SDSS-detected galaxies among which
1017 galaxies are also detected by the BATC photometry, corresponding to a
completeness of 90.7\%.

\section{Analysis of cluster galaxies with known spectroscopic redshifts  }
\subsection{Sample of Spectroscopically-Confirmed Member Galaxies}

For investigating the properties of galaxy cluster A119, 368 normal
galaxies with have known spectroscopic redshifts ($z_{sp}$) are
extracted in our viewing field from the NASA/IPAC Extragalactic
Database (NED). We cross-identified all these galaxies with the
BATC-detected sources.

In order to eliminate foreground and background galaxies, we apply a
standard iterative 3$\sigma$-clipping algorithm (\citealt{yahil97})
to the velocity distribution. For a galaxy cluster with complex
dynamics, the velocity distribution is expected to be non-gaussian.
Using the ROSTAT software, we can derive two resistant and robust
estimators, the biweight location (C$_{\rm BI}$) and scale (S$_{\rm
BI}$), analogous to the velocity mean and dispersion
(\citealt{beers90}). The galaxies with velocities between (C$_{\rm
BI}-3$C$_{\rm BI}$) and (C$_{\rm BI}+3$C$_{\rm BI}$) are selected as
member galaxies. After reaching the convergence, we achieve C$_{\rm
BI}$=13298$^{+93}_{-96}$ \kms and S$_{\rm BI}$=854$^{+80}_{-65}$
\kms. The errors of these two estimators correspond to $90\%$
confidence interval, and they are calculated by bootstrap resamplings
of 10,000 subsamples of the velocity data. There are 238 member
galaxies with 10736 \kms $< cz <$ 15860 \kms, and we refer to these
galaxies as ``sample I''. Based on 153 member galaxies of A119,
\citet{way97} derived C$_{\rm BI}$=13228$^{+103} _{-98}$ \kms and
S$_{\rm BI}$=778$^{+122}_{-88}$ \kms, slightly smaller than our
estimation. \citet{fabricant93} derived a mean radial velocity of
$13293\pm 101$ \kms and a velocity dispersion of $857^{+78}_{-61}$
\kms based on 80 member galaxies, which is in good agreement with our
statistics.

Fig.~2 shows the distribution of spectroscopic redshifts for those
368 galaxies. We take the NED-given cluster redshift of
$\bar{z_c}=0.0442$ for A119. For testing normality of velocity
distribution, we apply the Shapiro-Wilk W test to sample I, and
obtain the statistic $W=0.995$, corresponding a probability value of
$P=0.648$ which is much greater than the critical value $P=0.05$.
This indicates that the velocity distribution for sample I prove to
be consistent with Gaussian. The embedded panel of Fig.~2 shows the
histogram of radial velocities with a Gaussian fit. With the
velocities and positions of these member galaxies, mass of A119 can
be derived by applying virial theorem, assuming this cluster is well
virialized (\citealt{geller73,oegerle94}). We obtain a virial mass of
$8.86 \times 10^{14}M_{\odot}$. Table~2 lists the spectroscopic
redshifts of 238 member galaxies in sample I.

\begin{table}[t]
\centering
\begin{minipage}[]{125mm}
\caption[]{ Catalog of 238 Spectroscopically Confirmed Member
Galaxies in A119 }
\end{minipage}
\begin{tabular}{rcccc|rcccc}
\hline\noalign{\smallskip}
 No. & R.A. & Decl. & $z_{\rm sp}$ &  Ref.
 & No. & R.A.& Decl & $z_{\rm sp}$ & Ref. \\
\noalign{\smallskip} \hline\noalign{\smallskip}
1   &  00 56 21.0  &  -01 13 33  &  0.036755  &  (1) &   60  &  00 56 27.6 &  -01 23 15  &  0.042379   & (1)  \\
2   &  00 56 18.4  &  -01 08 04  &  0.036889  &  (2) &   61  &  00 56 16.2 &  -01 18 50  &  0.042416   & (1)  \\
3   &  00 56 24.7  &  -01 16 42  &  0.037232  &  (1) &   62  &  00 56 42.8 &  -01 13 26  &  0.042429   & (1)  \\
4   &  00 56 11.1  &  -01 19 42  &  0.037489  &  (1) &   63  &  00 56 20.3 &  -01 14 33  &  0.042476   & (1)  \\
5   &  00 56 12.9  &  -01 15 48  &  0.037783  &  (1) &   64  &  00 56 07.0 &  -01 28 28  &  0.042499   & (1)  \\
6   &  00 56 25.6  &  -01 15 45  &  0.038223  &  (3) &   65  &  00 56 01.9 &  -01 32 59  &  0.042563   & (2)  \\
7   &  00 56 29.2  &  -01 13 36  &  0.038403  &  (1) &   66  &  00 55 54.4 &  -01 03 50  &  0.042586   & (1)  \\
8   &  00 56 17.9  &  -01 15 43  &  0.038580  &  (1) &   67  &  00 55 50.7 &  -01 11 20  &  0.042613   & (1)  \\
9   &  00 55 51.5  &  -01 14 04  &  0.038588  &  (4) &   68  &  00 56 02.7 &  -01 20 04  &  0.042689   & (3)  \\
10  &  00 56 32.4  &  -01 11 15  &  0.038677  &  (2) &   69  &  00 55 30.5 &  -01 24 51  &  0.042763   & (2)  \\
11  &  00 57 03.0  &  -01 20 42  &  0.038680  &  (1) &   70  &  00 55 31.1 &  -01 13 24  &  0.042793   & (8)  \\
12  &  00 56 21.0  &  -01 10 36  &  0.039130  &  (2) &   71  &  00 55 39.9 &  -01 29 21  &  0.042806   & (1)  \\
13  &  00 56 17.8  &  -01 15 37  &  0.039154  &  (2) &   72  &  00 56 56.5 &  -00 56 37  &  0.042836   & (2)  \\
14  &  00 57 13.7  &  -01 00 44  &  0.039230  &  (2) &   73  &  00 56 17.1 &  -01 23 07  &  0.042840   & (2)  \\
15  &  00 56 18.0  &  -01 16 22  &  0.039254  &  (1) &   74  &  00 56 39.7 &  -01 28 29  &  0.042863   & (2)  \\
16  &  00 54 42.1  &  -01 29 22  &  0.039561  &  (5) &   75  &  00 55 45.9 &  -01 12 27  &  0.043023   & (1)  \\
17  &  00 55 28.5  &  -01 19 25  &  0.039747  &  (1) &   76  &  00 57 48.7 &  -01 00 15  &  0.043090   & (2)  \\
18  &  00 55 29.5  &  -01 07 58  &  0.039771  &  (2) &   77  &  00 56 06.0 &  -01 03 26  &  0.043123   & (2)  \\
19  &  00 55 46.8  &  -01 17 08  &  0.039948  &  (1) &   78  &  00 57 51.7 &  -01 08 10  &  0.043143   & (4)  \\
20  &  00 56 26.1  &  -01 09 25  &  0.040011  &  (1) &   79  &  00 55 47.8 &  -01 42 10  &  0.043223   & (2)  \\
21  &  00 54 38.7  &  -01 10 11  &  0.040066  &  (4) &   80  &  00 55 05.0 &  -01 15 09  &  0.043313   & (2)  \\
22  &  00 57 09.7  &  -01 22 49  &  0.040188  &  (2) &   81  &  00 57 14.6 &  -01 17 09  &  0.043370   & (2)  \\
23  &  00 55 38.2  &  -01 11 00  &  0.040355  &  (1) &   82  &  00 55 40.7 &  -01 18 44  &  0.043423   & (3)  \\
24  &  00 56 46.0  &  -01 32 39  &  0.040365  &  (2) &   83  &  00 57 26.3 &  -01 20 37  &  0.043450   & (2)  \\
25  &  00 56 18.6  &  -01 13 07  &  0.040365  &  (1) &   84  &  00 56 57.0 &  -01 23 20  &  0.043497   & (2)  \\
26  &  00 56 39.3  &  -01 04 37  &  0.040441  &  (1) &   85  &  00 56 04.0 &  -01 13 03  &  0.043543   & (4)  \\
27  &  00 55 16.2  &  -01 30 50  &  0.040581  &  (2) &   86  &  00 55 27.0 &  -01 13 23  &  0.043544   & (4)  \\
28  &  00 55 55.6  &  -01 14 50  &  0.040778  &  (2) &   87  &  00 56 25.9 &  -01 16 30  &  0.043583   & (2)  \\
29  &  00 56 51.5  &  -01 16 21  &  0.041002  &  (2) &   88  &  00 57 59.9 &  -01 25 09  &  0.043607   & (5)  \\
30  &  00 56 02.3  &  -01 29 46  &  0.041058  &  (1) &   89  &  00 56 47.1 &  -00 52 41  &  0.043617   & (2)  \\
31  &  00 55 35.6  &  -01 14 08  &  0.041116  &  (4) &   90  &  00 56 18.1 &  -01 14 31  &  0.043647   & (2)  \\
32  &  00 55 57.2  &  -01 30 50  &  0.041182  &  (1) &   91  &  00 54 29.8 &  -01 10 17  &  0.043655   & (4)  \\
33  &  00 55 51.2  &  -00 48 50  &  0.041194  &  (4) &   92  &  00 56 45.3 &  -00 56 27  &  0.043658   & (4)  \\
34  &  00 55 32.9  &  -01 11 33  &  0.041205  &  (1) &   93  &  00 56 19.7 &  -01 17 10  &  0.043727   & (1)  \\
35  &  00 56 13.6  &  -01 13 38  &  0.041235  &  (1) &   94  &  00 56 47.1 &  -01 16 57  &  0.043800   & (1)  \\
36  &  00 55 42.0  &  -01 03 31  &  0.041235  &  (1) &   95  &  00 56 12.8 &  -00 50 02  &  0.043805   & (4)  \\
37  &  00 57 10.7  &  -01 01 08  &  0.041298  &  (2) &   96  &  00 56 37.2 &  -01 32 24  &  0.043817   & (1)  \\
38  &  00 55 21.4  &  -01 21 08  &  0.041472  &  (2) &   97  &  00 56 32.2 &  -01 21 04  &  0.043847   & (1)  \\
39  &  00 55 55.0  &  -01 02 59  &  0.041569  &  (1) &   98  &  00 56 16.4 &  -01 32 36  &  0.043897   & (2)  \\
40  &  00 56 20.3  &  -01 15 02  &  0.041575  &  (1) &   99  &  00 57 11.6 &  -01 37 11  &  0.043934   & (8)  \\
41  &  00 55 25.2  &  -01 18 32  &  0.041585  &  (2) &   100 &  00 56 27.4 &  -00 47 33  &  0.043940   & (2)  \\
42  &  00 56 37.8  &  -01 20 41  &  0.041589  &  (1) &   101 &  00 55 52.2 &  -01 30 06  &  0.043957   & (2)  \\
43  &  00 54 41.0  &  -01 34 29  &  0.041625  &  (6) &   102 &  00 56 29.3 &  -01 20 25  &  0.044034   & (2)  \\
44  &  00 55 44.9  &  -01 12 57  &  0.041649  &  (4) &   103 &  00 56 41.6 &  -01 19 33  &  0.044080   & (1)  \\
45  &  00 56 47.4  &  -00 55 12  &  0.041686  &  (4) &   104 &  00 55 45.6 &  -01 03 17  &  0.044110   & (1)  \\
46  &  00 55 22.9  &  -01 12 40  &  0.041782  &  (2) &   105 &  00 56 38.5 &  -01 23 26  &  0.044117   & (2)  \\
47  &  00 55 39.0  &  -01 25 15  &  0.041842  &  (1) &   106 &  00 56 23.5 &  -00 59 13  &  0.044137   & (2)  \\
48  &  00 57 53.7  &  -00 48 53  &  0.041849  &  (6) &   107 &  00 55 16.9 &  -00 55 05  &  0.044179   & (4)  \\
49  &  00 56 10.7  &  -01 07 01  &  0.041856  &  (2) &   108 &  00 55 38.2 &  -01 16 46  &  0.044197   & (1)  \\
50  &  00 55 47.2  &  -01 13 53  &  0.041876  &  (1) &   109 &  00 57 02.0 &  -00 52 31  &  0.044238   & (4)  \\
51  &  00 56 22.2  &  -01 06 43  &  0.041956  &  (1) &   110 &  00 55 54.2 &  -00 55 21  &  0.044264   & 10)  \\
52  &  00 56 38.2  &  -01 17 53  &  0.041972  &  (1) &   111 &  00 55 12.9 &  -01 19 18  &  0.044274   & (2)  \\
53  &  00 56 46.4  &  -01 16 51  &  0.042060  &  (7) &   112 &  00 56 40.9 &  -01 02 04  &  0.044277   & (2)  \\
54  &  00 55 45.6  &  -01 05 03  &  0.042139  &  (4) &   113 &  00 56 01.7 &  -01 03 52  &  0.044284   & (8)  \\
55  &  00 54 38.8  &  -01 34 27  &  0.042216  &  (6) &   114 &  00 55 16.4 &  -01 21 05  &  0.044291   & (1)  \\
56  &  00 55 43.2  &  -01 02 01  &  0.042266  &  (2) &   115 &  00 56 35.7 &  -01 15 56  &  0.044354   & (1)  \\
57  &  00 56 30.7  &  -01 10 22  &  0.042273  &  (2) &   116 &  00 55 53.6 &  -00 54 04  &  0.044357   & (2)  \\
58  &  00 54 26.1  &  -00 55 42  &  0.042321  &  (4) &   117 &  00 55 05.5 &  -01 14 02  &  0.044407   & (2)  \\
59  &  00 56 31.5  &  -01 21 42  &  0.042359  &  (2) &   118 &  00 56 58.5 &  -01 15 25  &  0.044434   & (2)  \\

\noalign{\smallskip}   \hline\noalign{\smallskip}
\end{tabular}
\end{table}

\begin{table}
\setcounter{table}{1} \centering

\begin{minipage}{35mm}
\caption{\it --- Continued.}\end{minipage}\vspace{0pt}

\begin{tabular}{ r c c c c| r c c c c}
\hline\noalign{\smallskip} No. & R.A. & Decl. & $z_{\rm sp}$ &  Ref.
 & No. & R.A.& Decl & $z_{\rm sp}$ & Ref. \\
\noalign{\smallskip}\hline \noalign{\smallskip}
119 &  00 55 51.2  &  -01 18 23  &  0.044444  &  (8) &   179 &  00 55 19.1 &  -01 11 16  &  0.046149   & (1)  \\
120 &  00 56 16.1  &  -01 15 19  &  0.044464  &  (3) &   180 &  00 56 15.8 &  -01 26 48  &  0.046182   & (1)  \\
121 &  00 56 03.3  &  -00 49 41  &  0.044464  &  (2) &   181 &  00 56 21.8 &  -01 26 58  &  0.046205   & (2)  \\
122 &  00 58 02.1  &  -00 49 37  &  0.044503  &  (4) &   182 &  00 55 16.3 &  -01 24 16  &  0.046232   & (1)  \\
123 &  00 57 08.5  &  -01 29 39  &  0.044591  &  (8) &   183 &  00 56 07.1 &  -01 20 36  &  0.046272   & (6)  \\
124 &  00 55 57.9  &  -01 14 24  &  0.044597  &  (4) &   184 &  00 54 50.6 &  -01 28 45  &  0.046295   & (2)  \\
125 &  00 56 39.6  &  -01 04 43  &  0.044602  &  (4) &   185 &  00 56 25.1 &  -01 18 37  &  0.046299   & (1)  \\
126 &  00 55 59.6  &  -01 32 09  &  0.044624  &  (1) &   186 &  00 54 54.6 &  -01 20 27  &  0.046305   & (1)  \\
127 &  00 56 38.7  &  -01 21 06  &  0.044627  &  (1) &   187 &  00 56 56.7 &  -01 27 11  &  0.046319   & (1)  \\
128 &  00 56 00.0  &  -01 16 16  &  0.044637  &  (1) &   188 &  00 56 07.8 &  -01 25 47  &  0.046365   & (1)  \\
129 &  00 56 12.8  &  -01 16 10  &  0.044671  &  (2) &   189 &  00 56 25.6 &  -01 30 41  &  0.046432   & (1)  \\
130 &  00 55 55.1  &  -01 03 51  &  0.044694  &  (2) &   190 &  00 57 03.0 &  -00 52 25  &  0.046472   & (2)  \\
131 &  00 55 49.6  &  -01 04 54  &  0.044716  &  (4) &   191 &  00 56 27.7 &  -01 01 27  &  0.046525   & (2)  \\
132 &  00 56 18.1  &  -01 37 33  &  0.044771  &  (2) &   192 &  00 56 41.0 &  -01 18 26  &  0.046542   & (1)  \\
133 &  00 55 40.8  &  -00 59 50  &  0.044779  &  (4) &   193 &  00 57 02.0 &  -00 52 47  &  0.046585   & (6)  \\
134 &  00 57 07.8  &  -01 27 54  &  0.044781  &  (8) &   194 &  00 56 26.6 &  -01 20 53  &  0.046615   & (1)  \\
135 &  00 56 31.4  &  -01 36 59  &  0.044801  &  (2) &   195 &  00 55 03.6 &  -01 19 03  &  0.046689   & (1)  \\
136 &  00 55 51.6  &  -01 30 23  &  0.044814  &  (1) &   196 &  00 56 18.8 &  -01 28 12  &  0.046812   & (1)  \\
137 &  00 56 04.9  &  -01 08 09  &  0.044830  &  (4) &   197 &  00 56 44.5 &  -00 52 29  &  0.046842   & (2)  \\
138 &  00 56 53.2  &  -01 17 42  &  0.044858  &  (1) &   198 &  00 54 49.4 &  -01 23 19  &  0.046972   & (1)  \\
139 &  00 55 18.8  &  -01 16 38  &  0.044888  &  (3) &   199 &  00 56 17.6 &  -01 17 43  &  0.047176   & (1)  \\
140 &  00 56 39.3  &  -01 34 39  &  0.044931  &  (8) &   200 &  00 57 41.6 &  -01 18 44  &  0.047186   & (6)  \\
141 &  00 57 04.9  &  -00 55 09  &  0.044951  &  (2) &   201 &  00 55 58.6 &  -01 12 10  &  0.047206   & (1)  \\
142 &  00 56 14.3  &  -01 08 40  &  0.044961  &  (2) &   202 &  00 54 53.4 &  -01 34 35  &  0.047266   & (2)  \\
143 &  00 55 54.5  &  -00 55 14  &  0.044964  &  (2) &   203 &  00 56 11.2 &  -01 07 40  &  0.047299   & (1)  \\
144 &  00 55 57.3  &  -01 22 15  &  0.044971  &  (1) &   204 &  00 56 11.3 &  -01 31 35  &  0.047373   & (1)  \\
145 &  00 55 45.4  &  -01 23 59  &  0.045008  &  (2) &   205 &  00 55 18.5 &  -01 19 05  &  0.047549   & (1)  \\
146 &  00 57 34.9  &  -01 23 28  &  0.045031  &  (3) &   206 &  00 57 15.5 &  -00 49 31  &  0.047688   & (4)  \\
147 &  00 56 27.9  &  -01 26 07  &  0.045061  &  (2) &   207 &  00 56 33.7 &  -01 09 52  &  0.047746   & (1)  \\
148 &  00 56 23.0  &  -01 14 58  &  0.045118  &  (1) &   208 &  00 55 13.6 &  -01 04 34  &  0.047773   & (2)  \\
149 &  00 57 31.8  &  -00 55 40  &  0.045154  &  (2) &   209 &  00 56 51.6 &  -01 00 26  &  0.047796   & (4)  \\
150 &  00 56 00.7  &  -01 27 03  &  0.045181  &  (1) &   210 &  00 55 51.1 &  -01 09 53  &  0.047810   & (2)  \\
151 &  00 56 46.6  &  -01 35 47  &  0.045211  &  (5) &   211 &  00 56 52.7 &  -01 07 53  &  0.047880   & (4)  \\
152 &  00 56 13.3  &  -01 16 12  &  0.045318  &  (2) &   212 &  00 56 30.0 &  -01 05 13  &  0.047916   & (4)  \\
153 &  00 56 38.4  &  -01 07 34  &  0.045358  &  (2) &   213 &  00 56 15.0 &  -01 15 47  &  0.047936   & (1)  \\
154 &  00 55 55.0  &  -01 14 49  &  0.045368  &  (1) &   214 &  00 55 39.7 &  -00 52 36  &  0.047954   & (4)  \\
155 &  00 57 22.7  &  -00 49 19  &  0.045376  &  (4) &   215 &  00 56 57.8 &  -01 09 29  &  0.048092   & (4)  \\
156 &  00 55 17.9  &  -01 17 11  &  0.045421  &  (2) &   216 &  00 56 30.4 &  -01 32 02  &  0.048167   & (2)  \\
157 &  00 55 03.2  &  -01 15 57  &  0.045508  &  (2) &   217 &  00 56 26.8 &  -01 19 48  &  0.048383   & (1)  \\
158 &  00 55 53.3  &  -01 06 59  &  0.045520  &  (4) &   218 &  00 56 10.4 &  -01 08 25  &  0.048393   & (2)  \\
159 &  00 55 11.2  &  -00 48 37  &  0.045623  &  (4) &   219 &  00 56 11.9 &  -01 16 39  &  0.048523   & (1)  \\
160 &  00 56 16.2  &  -01 09 46  &  0.045635  &  (1) &   220 &  00 55 43.7 &  -01 19 46  &  0.048614   & (1)  \\
161 &  00 55 08.9  &  -01 02 47  &  0.045655  &  (2) &   221 &  00 56 13.4 &  -01 14 18  &  0.048662   & (4)  \\
162 &  00 56 15.4  &  -01 05 54  &  0.045748  &  (1) &   222 &  00 57 06.9 &  -01 23 51  &  0.048707   & (2)  \\
163 &  00 55 45.3  &  -01 19 28  &  0.045755  &  (2) &   223 &  00 55 19.2 &  -01 31 01  &  0.048707   & (1)  \\
164 &  00 56 26.5  &  -01 37 34  &  0.045765  &  (2) &   224 &  00 57 11.1 &  -01 22 49  &  0.048827   & (6)  \\
165 &  00 57 39.8  &  -00 53 33  &  0.045789  &  (4) &   225 &  00 57 00.3 &  -00 49 31  &  0.048850   & (2)  \\
166 &  00 56 52.7  &  -01 09 33  &  0.045812  &  (2) &   226 &  00 56 11.3 &  -01 31 53  &  0.048947   & (2)  \\
167 &  00 55 08.8  &  -01 23 48  &  0.045825  &  (8) &   227 &  00 55 59.1 &  -01 18 01  &  0.048977   & (2)  \\
168 &  00 56 10.2  &  -01 16 04  &  0.045838  &  (2) &   228 &  00 55 38.2 &  -01 34 47  &  0.049030   & (6)  \\
169 &  00 55 25.8  &  -01 23 02  &  0.045908  &  (1) &   229 &  00 56 58.3 &  -01 22 45  &  0.049101   & (1)  \\
170 &  00 56 43.2  &  -01 23 45  &  0.045958  &  (1) &   230 &  00 56 48.7 &  -01 29 32  &  0.049564   & (1)  \\
171 &  00 55 32.3  &  -01 12 40  &  0.045967  &  (4) &   231 &  00 56 22.8 &  -01 12 35  &  0.049918   & (2)  \\
172 &  00 55 59.2  &  -01 09 49  &  0.045993  &  (4) &   232 &  00 56 56.9 &  -01 12 43  &  0.049991   & (4)  \\
173 &  00 56 56.0  &  -00 59 48  &  0.046020  &  (4) &   233 &  00 56 39.0 &  -01 17 43  &  0.050155   & (5)  \\
174 &  00 57 06.4  &  -00 54 31  &  0.046032  &  (2) &   234 &  00 55 57.6 &  -01 17 26  &  0.050628   & (2)  \\
175 &  00 55 14.7  &  -00 49 20  &  0.046037  &  (4) &   235 &  00 56 39.1 &  -01 05 23  &  0.050902   & (1)  \\
176 &  00 55 30.8  &  -01 17 50  &  0.046052  &  (1) &   236 &  00 56 54.2 &  -01 15 25  &  0.051325   & (1)  \\
177 &  00 55 42.9  &  -01 11 46  &  0.046132  &  (1) &   237 &  00 56 34.5 &  -01 16 49  &  0.052399   & (1)  \\
178 &  00 56 11.3  &  -00 58 13  &  0.046139  &  (2) &   238 &  00 55 31.2 &  -01 11 33  &  0.052850   & (1)  \\

\noalign {\smallskip}   \hline\noalign{\smallskip}
\end{tabular}
\parbox{140mm}
{ References: {(1) \citet{cava09}; (2) \citet{smith04};
(3)\citet{wegner99}; (4) \citet{2003sdss}; (5) \citet{dale98}; (6)
\citet{rines03};  (7) \citet{kinman81}; (8) \citet{katgert98}} }
\end{table}


\begin{figure}[!ht]
\centering
\includegraphics[width=90mm,height=70mm]{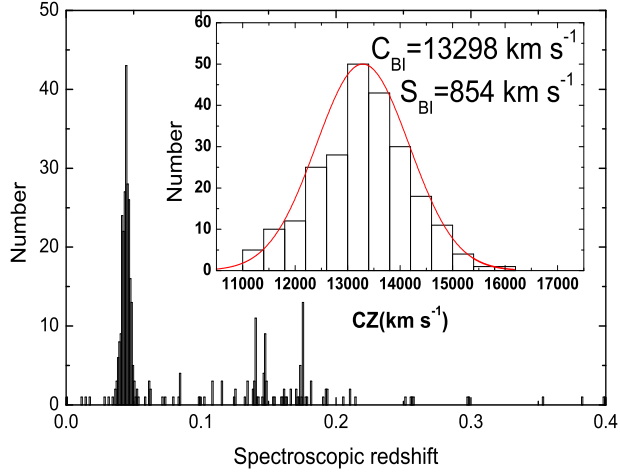}
\caption{\baselineskip 3.6mm The histogram of spectroscopic
redshifts between 0.0 and 0.4 for 368 galaxies detected by all multicolor surveys.
The embedded panel shows the histogram of radial velocity (\rm cz) distribution
of 238 galaxies in A119 in bins of 400 \kms.}
\label{fig2}
\end{figure}

\subsection{Spatial Distribution and Localized Velocity Structure}

Contour map of surface galaxy density is an intuitive tool in terms
of knowing existence of substructures in galaxy clusters. Left panel
of Fig.~3 shows contour maps of surface density with a smoothing
Gaussian window of $\sigma=2\arcmin$, superimposed on the projected
positions of 238 galaxies in sample I. As shown by the contour map,
A119 does not appear to be spherically asymmetric , and at least
three clumps can be found. It seems to agree with the X-ray
brightness map (see Fig.~1), which indicates the presence of
substructures in A119.

The clumps in spatial distribution might be due to projection effect.
For eliminating the ambiguity in substructure detection, we apply the
$\kappa$-test to sample I for quantifying the localized variation in
velocity distribution. The statistic $\kappa_n$ is introduced by
\citet{colless96} to quantify the local deviation on the scale of $n$
nearest neighbors:
$$\kappa_n = {\sum_{i=1}^N}-{\rm log}[P_{\rm KS}(D>D_{\rm obs})],$$
where $N$ is the number of member galaxies, and $[P_{\rm KS}(D>D_{\rm
obs})]$ represents probability of standard K-S statistic $D$ being
greater than observed value $D_{\rm obs}$. Thus a greater $\kappa_n$
means a greater probability that the local velocity distribution
differs from the overall distribution. The probability
P($\kappa_n>\kappa_n^{\rm obs}$) can be estimated by Monte Carlo
simulations by randomly shuffling velocities.
Table 4 gives the results of the $\kappa$-test for samples I and II
(sample II will be defined in the next section). For the 238 member
galaxies in sample I, the probability P($\kappa_n>\kappa_n^{\rm
obs}$) is found to be less than 15\% for a range of neighbor sizes
($6\leq n \leq 12$), indicating a probable substructure presence in
A119 .

Bubble plots in the case of 9 nearest neighbors are shown in the
right panel of Fig.~3.  Bubble size at the position of each galaxy is
proportional to $-$log[P$_{\rm KS}$(D$>$D$_{\rm obs}$)].
Consequently, the larger bubbles indicate a greater difference
between local and overall velocity distributions. For sample I, three
remarkable bubble clumps, called A, B, and C, can be found. The
southern 'clump' shown in the contour map (see the left panel) is
untrue, because no bubble clustering appears at the same location.
Big bunch of bubbles in central region obviously indicates an
anomalous velocity distribution at center.  For observing the
anomalous kinematics in bubble clumps, stripe densities of velocity
distributions are presented in Fig.~4 . For clumps A and C, their
velocity distributions have anomalous dispersions. Compared with the
dispersion of overall sample I (S$_{\rm BI}=854^{+80}_{-65}$ \kms),
clump A has a very large dispersion (S$_{\rm BI}=1282^{+205}_{-152}$
\kms), while clump C has significantly small dispersion (S$_{\rm
BI}=443^{+110}_{-342}$ \kms). Clump B is found to have smaller mean
velocity (C$_{\rm BI}=12733^{+710}_{-261}$ \kms). Keep in mind that
the errors of above biweight estimators correspond to a 90\%
confidence interval. Above biweight estimators in clumps differ from
those for whole sample at more than 3$\sigma$ significance level. It
is interesting that 85\% of the member galaxies with $cz < 11500$
\kms\ are found in clump A, which unambiguously points to a merging
along the line-of-sight direction in cluster core. Clump C seems to
be a compact group of galaxies with a bulk velocity of 12966 \kms.
The classic Kolmogorov-Smirnov (K-S) test shows the difference
for velocity distributions between each clump (A/B/C) and the overall
sample I are very significant, corresponding to the probabilities of 91.4\%, 
96\%, and 95.2\%, respectively. This means three clumps are probably real substructures.

\begin{figure}[!h]
\centering
\includegraphics[width=70mm,angle=-90]{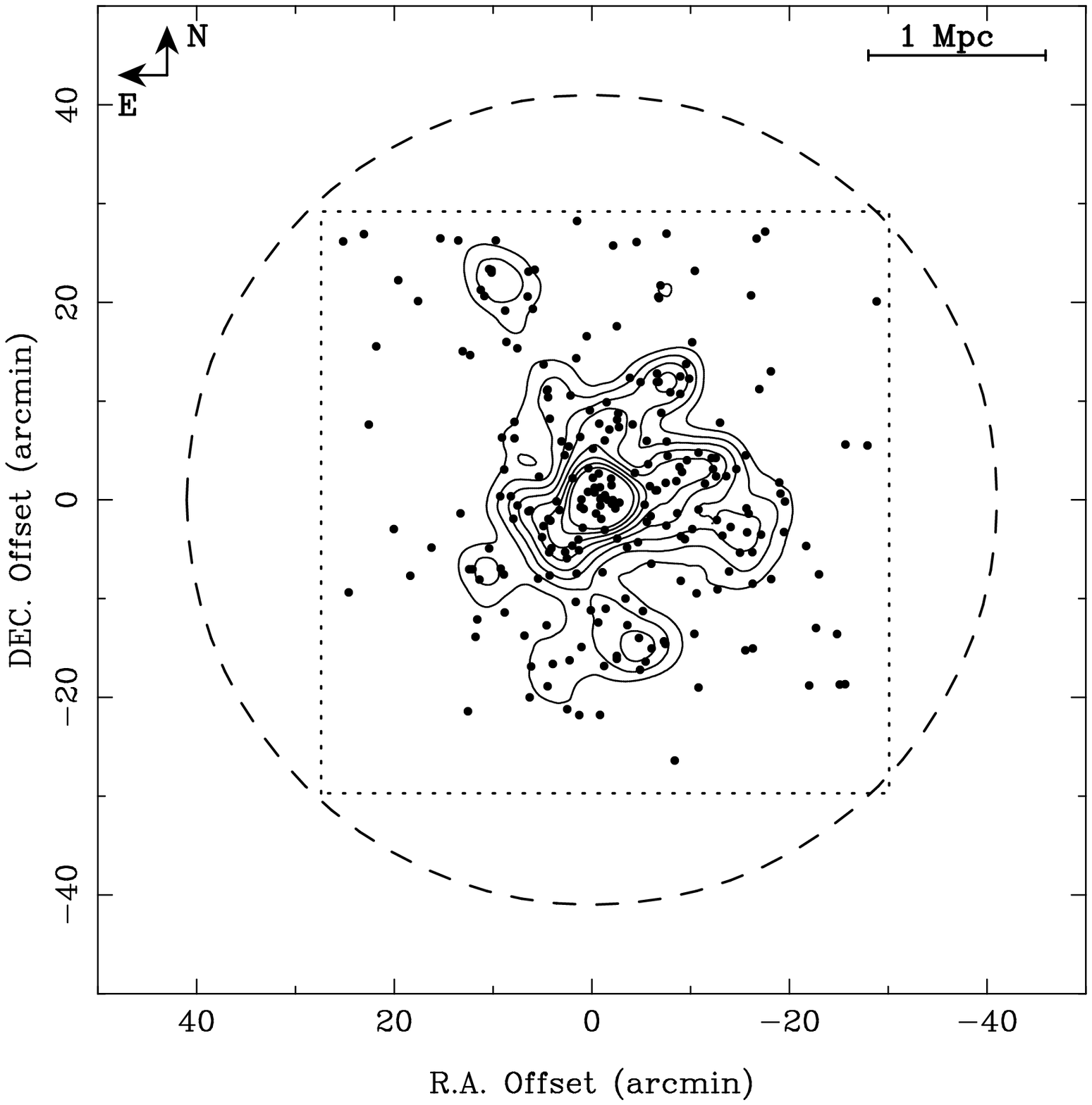}
\includegraphics[width=70mm,angle=-90]{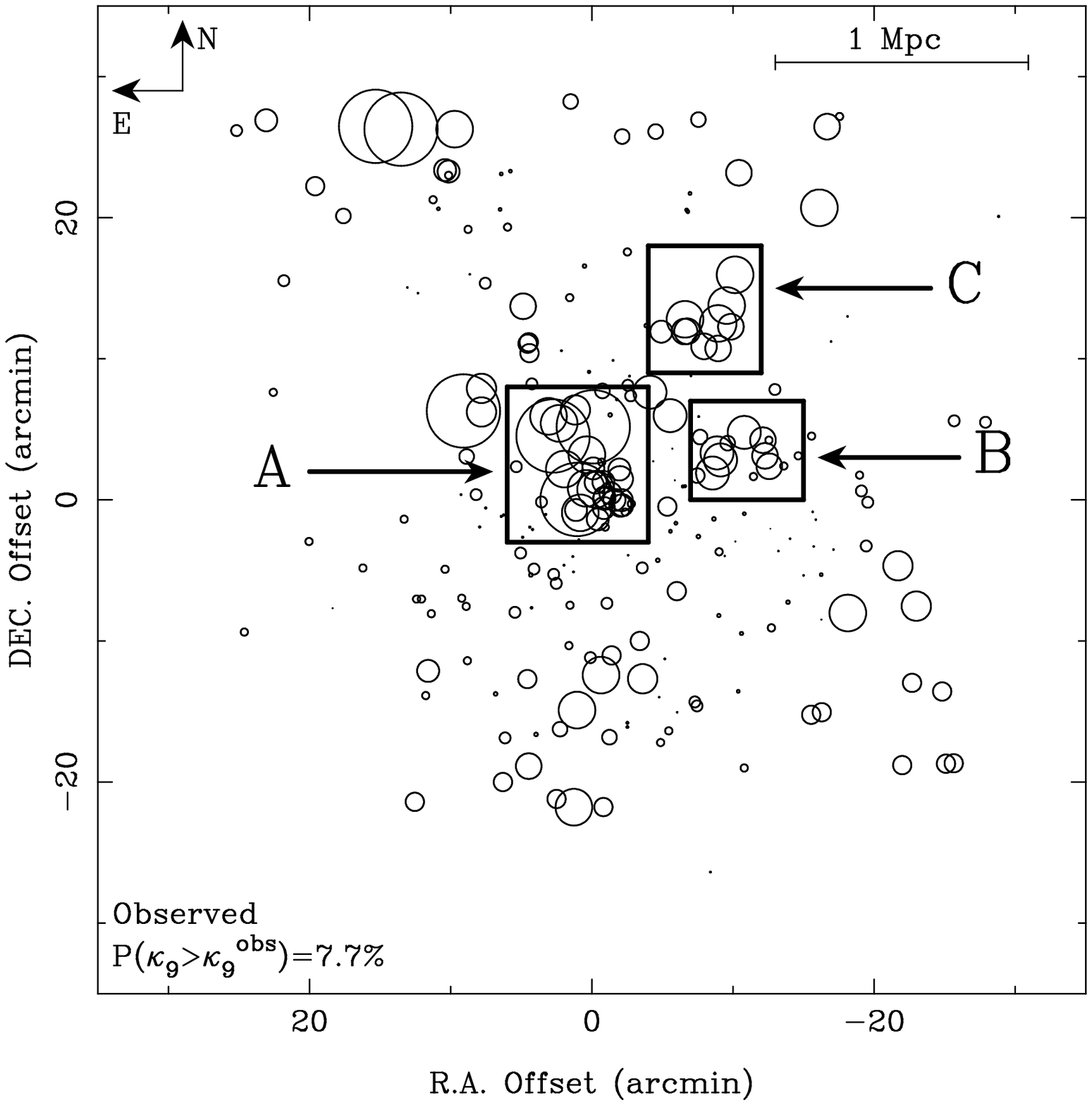}
\caption{ \baselineskip 3.6mm Left: Spatial distribution for 238
spectroscopic member galaxies of A119 in sample I(denoted by
filled circles). The contour map of the surface density use
the smoothing Gaussian window $\sigma=2\arcmin$ .
Contour map is superposed with the surface
density levels 0.11, 0.16, 0.21, 0.26, 0.31, 0.36, 0.41
 and 0.46 arcmin$^{-2}$. The dashed big circles show a typical
 region of rich clusters with a radius of $R \approx 1.5 h^{-1}$
 Mpc of the cluster center.
The region within dotted line represents BATC field.
Right:Bubble plot showing the localized
variation for groups of the 9 nearest neighbors for 238 member
galaxies in sample I.
\label{fig3}}
\end{figure}

\begin{figure}[!h]
\centering
\includegraphics[width=90mm,height=70mm]{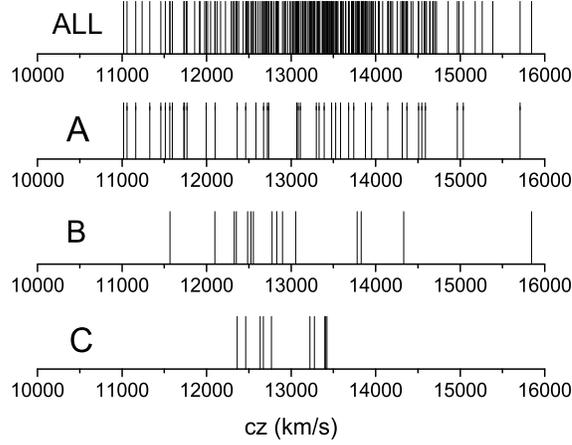}
\caption{Stripe density plot of velocity of the spectroscopically
confirmed galaxies in whole cluster, clumps A, B and C, respectively.
\label{fig4}}
\end{figure}

\section{Photometric redshft technique and selection of faint member}

\subsection{Membership Selection by Photometric Redshift Technique}

Although optical spectroscopy is a straightforward approach for the
cluster membership determination, spectroscopy of faint galaxies
remains a rather daunting task. Photometric redshift technique plays
a crucial role in finding faint galaxies based on their spectral
energy distributions (SEDs) (\citealt{pello99,brunner00}).

Technique of photometric redshift has been applied extensively in
deep photometric survey with large-field detectors
(\citealt{lan96,arn99, fur00}). A standard SED-fitting code, called
HYPERZ (\citealt{bol00}), is adopted for estimating the photometric
redshifts. The procedure has been adapted especially for the BATC
multicolor photometric system (\citealt{yuan01,xia02,zhou03b}). The
SED templates for normal galaxies are generated through convolving
the galaxy spectra in template library GISSEL98 (Galaxy Isochore
Synthesis Spectral Evolution Library; \citet{bru93}) with
transmission curves of the BATC filters. For a given source, the
photometric redshift, $z_{ph}$, corresponds to the best fit (in the
$\chi^{2}$-sense) between photometric SED and template SED. The
reddening law of Milky Way (\citealt{allen76}) is adopted for dust
extinction, and the A$_{V}$ is flexible in a range from 0.0 to 0.2.

A cross-identification between SDSS photometric catalog and the
BATC-detected sources, 1376 galaxies in our viewing field are
extracted, including 1008 faint galaxies without spectroscopic
redshifts. As a test, we firstly let the {\it z$_{ph}$} vary in a
wide range from 0.0 to 1.0, with steps of 0.01. Only small number of
galaxies are found to have $z_{ph}>0.4$. We search the photometric
redshifts for 1376 galaxies brighter than $i_{\rm BATC}=19.5$ in a
range from 0.0 to 0.4, with a step of 0.001.

To appraise the precision of our $z_{ph}$ estimation, a comparison
between $z_{ph}$ and $z_{sp}$ values for the galaxies brighter than
$i_{\rm BATC}$=17.0 is given in fig.~5. The dashed lines indicate an
average redshift deviation of 0.0103, and error bars of {\it
z$_{ph}$} correspond to 68\% confidence level in photometric redshift
determination. Our {\it z$_{ph}$} estimate is basically in accordance
with the {\it z$_{sp}$} values. In Fig.5, there are five galaxies
whose photo-z values are significantly deviated from the spectroscopic
redshifts. We check images and SEDs of these galaxies, and find that
the SEDs of some galaxies suffer from some satellite contamination
within the aperture, and some galaxies locate at the edge of BATC field,
which result in false SEDs.

An iterative 2$\sigma$-clipping algorithm is applied to the $z_{ph}$
values of 238 member galaxies in sample I. We achieve C$_{\rm BI}$=
0.048$^{+0.01}_{-0.01}$ and S$_{\rm BI}$=0.09$^{+0.01}_{-0.01}$ for
{\it z$_{ph}$} estimate. Statistically, 137 of 144 member galaxies
(about 95\%) brighter than $i_{BATC}$=17.0 are found to have their
photometric redshifts within $\pm 2S_{\rm BI}$ deviation,
corresponding to a $z_{ph}$ range from 0.030 to 0.066. Even for all
238 member galaxies in sample I, 192 galaxies (about 81\%) are found
to have $0.030<z_{ph}<0.066$. The $z_{ph}$ range is taken as a
selection criterion of faint member candidates. As a result, 144
galaxies with $0.030<z_{ph}<0.066$ are regarded as new member
candidates of A119. The distributions of photometric redshift for
sample I and new member candidates are shown in Fig.~6. The dashed
lines denote the $z_{ph}$ range of selection criterion.

\begin{figure}[!ht]
\begin{minipage}[t]{0.5\linewidth}
\centering
\includegraphics[width=70mm,height=70mm]{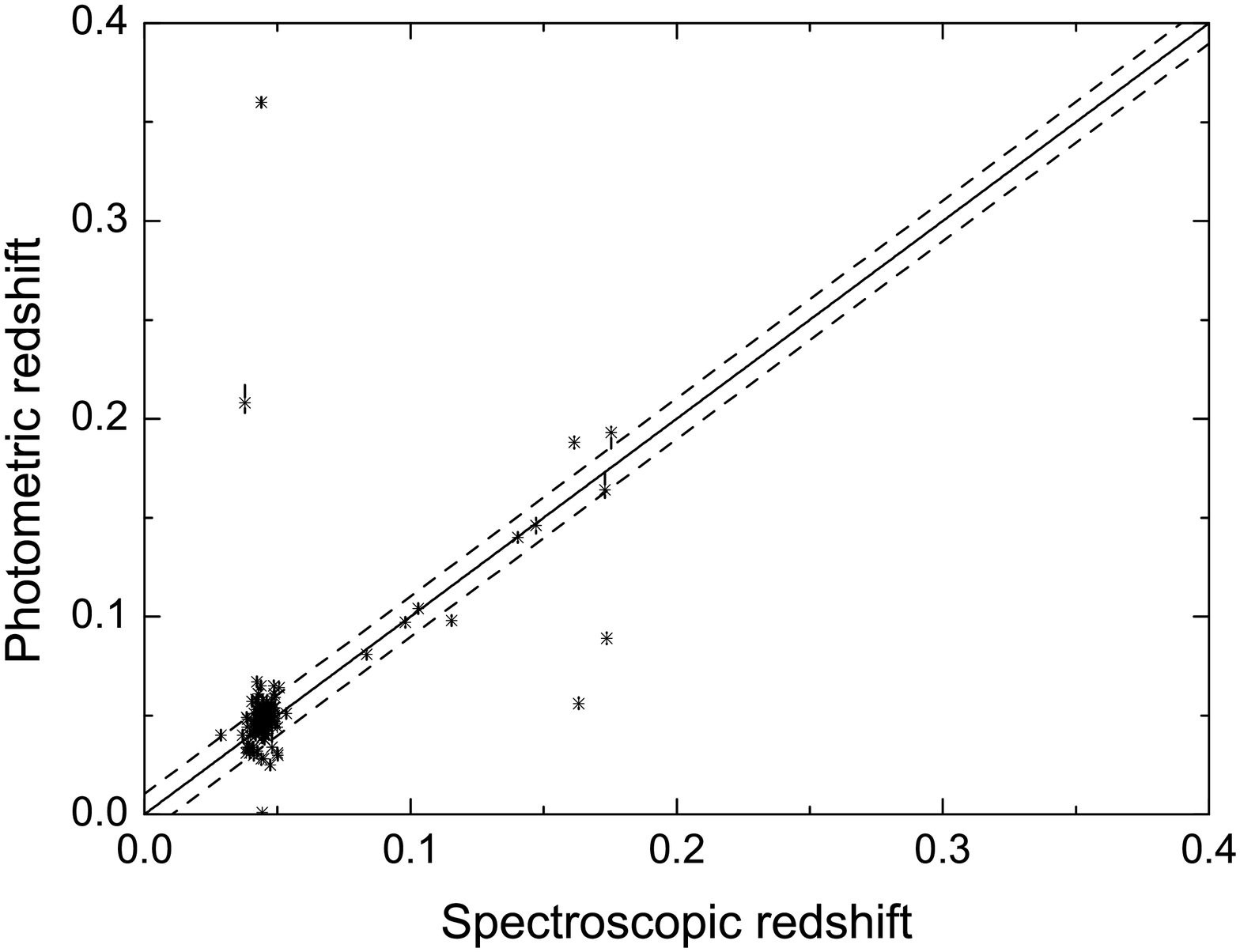}
\caption{Comparison between photometric redshift ({\it z$_{ph}$}) and
spectroscopic redshift ({\it z$_{sp}$}) for 157 galaxies brighter
than $i_{\rm BATC}=17.0$ with known spectroscopic redshifts
 in A119. The solid line corresponds to  $z_{sp}
 = z_{ph}$, and dashed lines  indicate an average
 deviation of 0.0103.}
\label{fig5}
\end{minipage}
\begin{minipage}[t]{0.5\linewidth}%
\centering
\includegraphics[width=70mm,height=70mm]{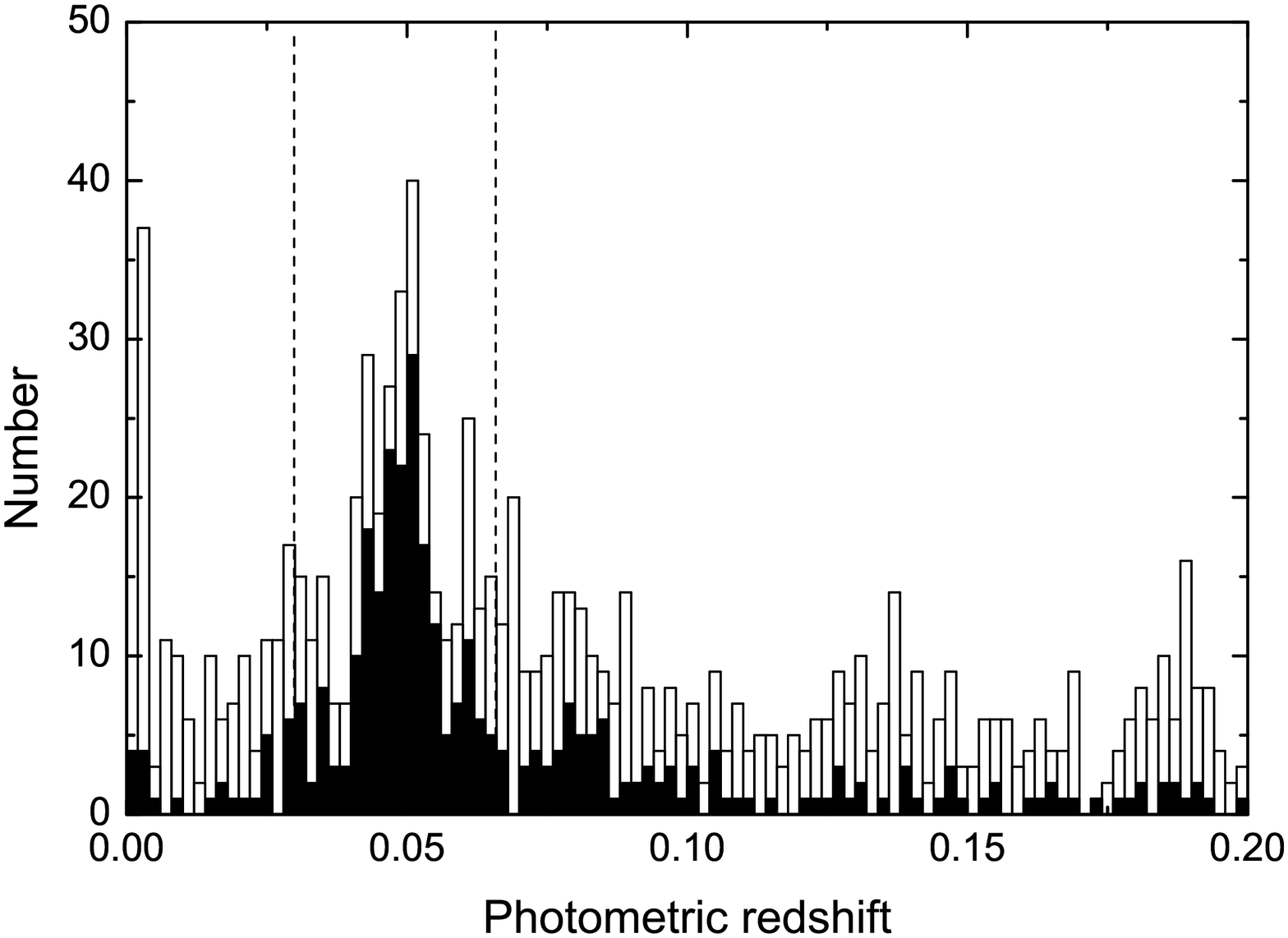}
\caption {The distribution of  photometric redshifts, with  a bin
size of 0.002. The black histogram shows {\it z$_{ph}$} 238 galaxies
in sample I. The dashed lines represent the {\it z$_{ph}$} range of
selection criterion. \label{fig6}}
\end{minipage}
\end{figure}

\subsection{Color-Magnitude Correlation}
A universal correlation between color and absolute magnitude for
early-type galaxies, called CM relation, has been commonly found in
rich galaxy clusters(\citealt[and references therein]{bower92}).
Brighter early-type galaxies tend to be redder. The SDSS photometric
catalog provides a parameter, \texttt{fracDeV}, defined as fraction
of the brightness contribution by de Vaucouleurs component. We take
the galaxies with \texttt{fracDeV} $>$ 0.5 as early-types, and the
remaining as late-type galaxies. The CM relation can be used to
constrain the membership selection of early-type galaxies
(\citealt{yuan01}).

Fig.~7 shows the relation between color index $b-h$ and $h$ magnitude
for member galaxies in sample I and newly-selected member candidates.
A linear fitting is performed for the 181 early-type galaxies in
sample I: $b-h=-0.20(\pm0.01)h+5.44(\pm0.19)$, and dashed lines
denote $\pm 3 \sigma$ deviation. A majority of early-type member
galaxies in sample I follows a tight color-magnitude correlation.
However, 27 early-type candidates of member galaxies are scattered
beyond the $3\sigma$ deviation of intercept, and they are removed
from our candidate list.

Finally, a list of 355 member galaxies is obtained by combining the
238 spectroscopically confirmed member galaxies in sample I
(including 181 early-type galaxies and 57 late-type galaxies) with
the 117 newly selected member galaxies (including 36 early-type
galaxies and 81 late-type galaxies), to which we refer as sample II.
SED information, the SDSS morphological parameter `\texttt{fracDeV}',
and photometric redshifts for the 117 newly selected member galaxies
are catalogued in Table~3. The magnitude of 99.00 means non-detection
in the specified band.

For estimating the percentage of object blending due to the seeing effect, 
a cross-identification of the 117 new candidates of member galaxies is 
performed with the photometric catalog of the SDSS galaxies. A searching 
circle with a radius of 4.\arcsec5 centered at BATC-detected galaxies is 
adopted, and 3 galaxies are found to have more than one counterpart within 
the searching region, corresponding to a small probability (2.6\%) of object 
blending.

\begin{figure}[!ht]
\centering
\includegraphics[width=70mm,height=70mm,angle=-90]{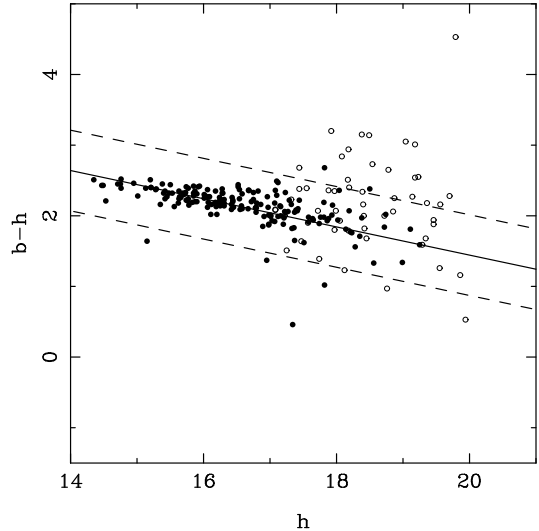}
\caption{Color-magnitude relation for early-type galaxies in A119.
Early-type member galaxies with known spectroscopic redshifts are
denoted by {\it filled circles}, and newly selected early-type member
candidates are denoted by {\it open circles}. The solid line shows
the linear fit for 181 early-type galaxies in sample I. The dashed
lines correspond to the $\pm3\sigma$ deviation of intercept.
\label{fig7}}
\end{figure}

\begin{table}[t]
\centering \begin{minipage}{125mm}

\caption{Catalog of 117 Newly-selected Candidates of Member
Galaxies in A119}\end{minipage}

\scriptsize \tabcolsep 0.70mm
\renewcommand\arraystretch{1.1}
\begin{tabular}{rccccccccccccccccccc}
\hline\noalign{\smallskip} {No.}&{R.A.} &{Decl.} & {$z_{\rm ph}$} &
{fracDeV} & {$a$} & {$b$} & {$c$} & {$d$} & {$e$} & {$f$} & {$g$} &
{$h$} & {$i$} & {$j$} & {$k$} & {$m$}
& {$n$} & {$o$} & {$p$} \\
\noalign{\smallskip}\hline\noalign{\smallskip}
1   & 0 57 43.29 &  -1 30 40.00  &  0.030 &  0.080 &  19.50&  20.13&  18.84&  99.00&  19.32&  19.49&  19.18&  19.12&  18.66&  18.80&  18.55&  18.80&  18.48&  18.64&  18.52 \\
2   & 0 56 17.74 &  -1 19 30.50  &  0.030 &  0.000 &  99.00&  21.65&  19.15&  20.69&  20.59&  20.24&  99.00&  21.08&  19.48&  18.93&  20.40&  19.60&  19.38&  20.46&  99.00 \\
3   & 0 54 51.55 &  -1 13 24.20  &  0.030 &  0.376 &  21.65&  21.02&  19.58&  19.38&  18.76&  18.66&  18.34&  18.27&  18.11&  18.02&  17.86&  17.85&  17.68&  17.67&  17.88 \\
4   & 0 56 48.50 &  -0 54 10.50  &  0.030 &  0.474 &  21.83&  21.73&  21.07&  20.11&  20.38&  20.24&  19.31&  19.26&  19.19&  19.43&  19.19&  19.22&  18.71&  19.18&  18.26 \\
5   & 0 57 35.82 &  -1 23 13.30  &  0.030 &  1.000 &  21.27&  19.98&  19.44&  19.06&  18.50&  18.47&  18.33&  18.05&  17.75&  17.62&  17.52&  17.49&  17.22&  17.08&  17.59 \\
6   & 0 58 02.94 &  -1 27 55.20  &  0.031 &  0.504 &  19.41&  19.01&  18.74&  99.00&  18.69&  18.44&  18.70&  99.00&  18.28&  18.37&  18.46&  99.00&  18.52&  18.44&  18.26 \\
7   & 0 57 09.83 &  -1 11 31.10  &  0.031 &  1.000 &  21.62&  20.13&  19.84&  18.91&  19.09&  18.79&  18.32&  18.45&  18.11&  18.20&  18.00&  17.84&  17.61&  17.54&  17.36 \\
8   & 0 56 31.40 &  -1 33 42.40  &  0.032 &  0.082 &  20.66&  19.97&  19.51&  99.00&  18.89&  18.80&  18.31&  18.49&  18.13&  18.10&  18.00&  17.89&  17.68&  17.79&  17.65 \\
9   & 0 55 59.26 &  -1 16 11.90  &  0.032 &  0.000 &  21.85&  20.46&  20.60&  20.52&  20.24&  19.73&  20.68&  19.53&  19.30&  99.00&  19.17&  19.45&  18.85&  18.76&  18.53 \\
10  & 0 56 25.61 &  -1 10 47.20  &  0.032 &  0.333 &  20.98&  20.22&  19.73&  19.56&  18.88&  18.86&  18.46&  18.55&  18.28&  18.39&  18.14&  18.19&  18.03&  17.86&  17.53 \\
11  & 0 55 41.29 &  -0 56 30.70  &  0.032 &  0.358 &  19.49&  18.87&  18.62&  18.47&  18.22&  18.23&  18.20&  18.03&  17.81&  17.99&  17.86&  18.00&  17.69&  17.56&  17.45 \\
12  & 0 56 30.90 &  -1 09 13.40  &  0.032 &  0.295 &  21.58&  20.39&  20.09&  19.78&  19.53&  19.28&  18.88&  18.76&  18.59&  18.64&  18.50&  18.45&  18.30&  18.26&  17.69 \\
13  & 0 56 16.92 &  -1 32 28.70  &  0.033 &  0.384 &  20.61&  19.51&  19.23&  99.00&  18.43&  18.35&  18.11&  17.97&  17.88&  17.88&  17.63&  17.59&  17.41&  17.35&  17.73 \\
14  & 0 55 45.99 &  -1 24 07.90  &  0.033 &  0.417 &  21.75&  20.13&  20.33&  19.99&  19.15&  99.00&  19.32&  18.93&  18.60&  18.68&  18.40&  18.44&  18.26&  18.12&  18.42 \\
15  & 0 57 04.88 &  -1 10 59.90  &  0.033 &  0.275 &  20.60&  19.21&  18.73&  18.57&  17.80&  18.00&  17.80&  17.86&  17.67&  17.57&  17.42&  17.41&  17.02&  17.01&  99.00 \\
16  & 0 55 26.52 &  -1 37 50.50  &  0.034 &  0.000 &  19.57&  19.19&  18.52&  99.00&  18.05&  17.98&  17.75&  17.62&  17.54&  17.45&  17.57&  17.39&  17.37&  17.23&  17.43 \\
17  & 0 56 57.82 &  -1 33 16.80  &  0.034 &  0.011 &  20.25&  19.82&  19.08&  99.00&  18.66&  18.45&  18.41&  18.07&  18.07&  17.93&  17.91&  17.84&  17.62&  17.64&  17.68 \\
18  & 0 56 14.31 &  -1 15 16.80  &  0.034 &  0.972 &  22.71&  99.00&  19.49&  19.16&  18.37&  18.22&  18.39&  17.93&  17.66&  17.82&  17.37&  17.26&  17.12&  16.97&  17.31 \\
19  & 0 56 28.79 &  -1 20 30.50  &  0.035 &  0.179 &  24.51&  20.77&  20.23&  20.09&  19.37&  99.00&  19.06&  19.10&  18.61&  18.94&  18.74&  18.55&  17.98&  18.27&  18.50 \\
20  & 0 56 28.82 &  -1 36 43.80  &  0.035 &  0.000 &  21.50&  20.04&  20.44&  99.00&  19.62&  19.70&  19.61&  19.43&  19.11&  19.19&  19.01&  19.22&  18.67&  18.68&  18.47 \\
21  & 0 55 20.59 &  -1 12 31.60  &  0.036 &  0.154 &  21.75&  20.10&  20.10&  19.56&  19.34&  19.06&  18.71&  18.82&  18.55&  18.59&  99.00&  18.51&  18.00&  17.95&  18.60 \\
22  & 0 55 38.94 &  -1 09 46.80  &  0.036 &  0.030 &  21.07&  21.41&  20.31&  20.10&  19.50&  19.43&  19.44&  18.97&  18.99&  19.52&  99.00&  19.09&  18.60&  18.59&  18.50 \\
23  & 0 56 49.25 &  -1 01 15.80  &  0.036 &  0.076 &  21.84&  20.51&  20.21&  19.86&  19.63&  99.00&  19.03&  19.22&  18.74&  18.49&  18.77&  18.56&  18.29&  18.32&  18.30 \\
24  & 0 57 08.77 &  -0 56 48.00  &  0.036 &  0.393 &  22.08&  21.28&  20.20&  19.80&  19.48&  19.46&  18.82&  18.84&  18.67&  18.62&  18.62&  18.72&  18.20&  18.54&  18.45 \\
25  & 0 55 52.56 &  -1 21 31.60  &  0.038 &  0.725 &  20.59&  20.32&  19.40&  18.98&  18.61&  18.52&  18.44&  17.97&  17.89&  18.02&  17.87&  17.65&  17.56&  17.32&  17.95 \\
26  & 0 57 21.38 &  -1 16 39.80  &  0.038 &  0.129 &  99.00&  22.17&  21.12&  20.32&  20.01&  20.05&  19.57&  19.71&  19.48&  19.15&  19.10&  19.23&  18.87&  18.92&  20.79 \\
27  & 0 55 56.44 &  -1 36 46.80  &  0.039 &  0.209 &  19.98&  19.20&  18.61&  18.14&  17.78&  17.69&  17.36&  17.17&  17.06&  16.95&  16.94&  16.79&  16.66&  16.62&  99.00 \\
28  & 0 55 19.23 &  -1 16 29.10  &  0.039 &  0.000 &  19.60&  18.45&  17.84&  17.39&  16.89&  16.85&  16.64&  16.43&  16.23&  16.10&  16.09&  15.85&  15.65&  15.52&  15.50 \\
29  & 0 55 28.43 &  -1 41 13.90  &  0.040 &  0.274 &  21.55&  20.00&  19.52&  99.00&  18.79&  18.66&  99.00&  18.12&  18.05&  18.05&  17.99&  17.78&  17.57&  17.57&  17.69 \\
30  & 0 55 44.12 &  -1 29  2.10  &  0.040 &  0.248 &  21.97&  21.15&  20.34&  99.00&  19.68&  19.32&  19.21&  18.72&  18.80&  18.61&  18.66&  18.52&  18.41&  18.18&  18.76 \\
31  & 0 55 10.98 &  -1 04 51.30  &  0.040 &  0.157 &  20.19&  19.59&  19.36&  19.25&  19.11&  18.94&  19.27&  18.80&  18.90&  18.50&  19.01&  18.81&  18.91&  18.79&  99.00 \\
32  & 0 56 05.55 &  -1 03 22.90  &  0.040 &  0.489 &  20.30&  19.38&  18.64&  18.11&  17.75&  17.57&  17.34&  17.19&  16.98&  16.94&  16.85&  16.69&  16.63&  16.42&  99.00 \\
33  & 0 56 57.42 &  -0 57 34.70  &  0.040 &  0.370 &  20.64&  19.90&  19.33&  18.94&  18.47&  18.37&  18.04&  17.90&  17.72&  17.74&  17.55&  17.50&  17.37&  17.30&  17.03 \\
34  & 0 55 25.87 &  -0 57 01.20  &  0.040 &  0.950 &  21.11&  20.91&  20.09&  19.90&  19.64&  19.18&  19.05&  18.85&  18.68&  18.58&  99.00&  18.61&  18.24&  18.34&  17.83 \\
35  & 0 55 37.82 &  -0 56 01.00  &  0.040 &  0.000 &  20.56&  19.97&  19.76&  19.65&  18.80&  19.08&  18.93&  18.82&  18.85&  18.20&  99.00&  18.38&  17.95&  17.78&  17.99 \\
36  & 0 55 19.42 &  -0 48 38.50  &  0.040 &  0.094 &  21.31&  21.56&  19.85&  19.40&  19.12&  19.02&  18.53&  18.78&  18.56&  18.56&  99.00&  18.53&  99.00&  18.24&  18.07 \\
37  & 0 56 38.89 &  -1 00 48.00  &  0.041 &  0.986 &  22.04&  99.00&  21.49&  99.00&  21.46&  20.29&  19.10&  19.33&  19.44&  99.00&  19.78&  19.14&  18.41&  18.69&  17.57 \\
38  & 0 55 14.03 &  -1 04 08.20  &  0.042 &  0.317 &  99.00&  20.56&  20.32&  19.85&  19.59&  19.40&  18.92&  18.88&  18.69&  18.50&  99.00&  18.12&  17.85&  17.96&  17.72 \\
39  & 0 57 23.13 &  -1 43 33.10  &  0.042 &  0.502 &  21.49&  20.88&  21.09&  99.00&  19.81&  19.91&  19.11&  19.29&  19.41&  19.19&  18.91&  99.00&  18.48&  18.70&  18.32 \\
40  & 0 54 39.14 &  -1 36 02.00  &  0.042 &  0.589 &  19.45&  19.11&  18.39&  17.97&  17.87&  17.81&  17.57&  17.47&  17.35&  17.29&  17.34&  17.21&  17.13&  17.02&  17.09 \\
41  & 0 55 23.45 &  -1 17 22.00  &  0.042 &  0.578 &  21.11&  21.40&  20.34&  19.88&  19.84&  99.00&  18.85&  19.46&  18.80&  18.90&  19.29&  18.88&  18.72&  18.61&  20.00 \\
42  & 0 58 03.53 &  -1 40 35.80  &  0.042 &  0.000 &  21.85&  99.00&  20.64&  99.00&  20.02&  19.84&  19.86&  99.00&  19.23&  19.31&  19.31&  99.00&  18.64&  19.04&  18.93 \\
43  & 0 55 58.89 &  -1 17 49.50  &  0.043 &  0.000 &  21.96&  20.36&  20.32&  19.88&  19.53&  19.36&  19.47&  19.78&  18.86&  19.15&  19.11&  18.73&  18.44&  18.39&  17.78 \\
44  & 0 55 17.89 &  -1 17 10.30  &  0.043 &  0.631 &  19.82&  19.04&  18.22&  17.77&  17.24&  17.12&  16.86&  16.68&  16.49&  16.40&  16.34&  16.19&  16.07&  15.98&  15.96 \\
45  & 0 56 59.72 &  -0 51 02.80  &  0.043 &  0.246 &  21.01&  20.32&  20.33&  20.38&  19.23&  19.46&  19.34&  19.09&  19.16&  19.03&  18.98&  18.87&  18.43&  18.35&  18.65 \\
46  & 0 55 05.88 &  -1 39 30.20  &  0.044 &  0.259 &  21.32&  20.34&  20.32&  99.00&  19.57&  19.81&  19.94&  19.23&  19.28&  19.22&  19.13&  18.91&  18.76&  18.76&  18.23 \\
47  & 0 55 13.98 &  -1 07 30.20  &  0.044 &  0.441 &  21.59&  20.81&  20.12&  19.57&  19.28&  19.09&  18.72&  18.72&  18.55&  18.63&  18.60&  18.43&  18.28&  18.01&  19.91 \\
48  & 0 56 29.41 &  -1 11 41.30  &  0.045 &  0.139 &  20.68&  20.45&  19.71&  19.58&  18.92&  18.90&  18.34&  18.43&  18.25&  99.00&  18.03&  17.92&  17.86&  17.69&  17.59 \\
49  & 0 56 27.08 &  -1 20 53.90  &  0.045 &  0.898 &  20.23&  19.16&  18.63&  18.00&  17.66&  99.00&  17.20&  17.08&  16.86&  16.81&  99.00&  16.55&  16.45&  16.33&  16.26 \\
50  & 0 54 39.17 &  -1 34 32.50  &  0.045 &  1.000 &  19.95&  19.17&  18.45&  99.00&  17.53&  17.48&  17.18&  17.08&  16.84&  16.72&  16.74&  16.53&  16.44&  16.29&  16.46 \\
51  & 0 56 07.93 &  -1 00 52.20  &  0.046 &  0.350 &  21.41&  20.11&  20.22&  19.60&  19.25&  19.14&  19.00&  18.75&  18.45&  18.30&  18.06&  18.03&  17.85&  17.75&  99.00 \\
52  & 0 55 09.54 &  -1 03 03.40  &  0.046 &  0.000 &  21.71&  22.22&  20.03&  20.17&  18.87&  18.98&  18.83&  18.77&  18.54&  18.33&  18.53&  18.25&  18.24&  18.09&  18.42 \\
53  & 0 55 03.89 &  -1 22 40.20  &  0.047 &  0.586 &  21.20&  20.56&  19.73&  19.55&  18.96&  18.89&  18.74&  18.40&  18.22&  18.12&  18.39&  18.07&  17.82&  17.94&  17.87 \\
54  & 0 56 46.26 &  -1 36 13.80  &  0.048 &  1.000 &  20.43&  19.82&  18.94&  99.00&  18.06&  17.90&  17.64&  17.44&  17.31&  17.30&  17.14&  17.06&  16.94&  16.84&  16.73 \\
55  & 0 55 44.80 &  -1 34 49.80  &  0.048 &  0.933 &  20.53&  19.94&  19.09&  99.00&  18.13&  18.02&  17.69&  17.55&  17.42&  17.32&  17.29&  17.08&  16.91&  16.95&  16.80 \\
56  & 0 55 50.13 &  -1 28 38.10  &  0.048 &  0.000 &  20.92&  20.52&  20.13&  20.20&  19.69&  19.78&  19.13&  18.81&  19.25&  19.27&  19.69&  19.15&  19.07&  19.33&  99.00 \\
57  & 0 55 17.55 &  -1 16 47.00  &  0.048 &  0.336 &  22.25&  21.60&  19.81&  19.37&  18.78&  18.63&  18.31&  18.07&  17.77&  17.76&  17.56&  17.37&  17.27&  17.19&  16.96 \\
58  & 0 56 17.98 &  -1 15 00.90  &  0.048 &  0.000 &  22.40&  21.16&  19.87&  19.08&  18.68&  18.49&  17.96&  18.05&  17.79&  17.97&  17.58&  17.39&  17.34&  17.20&  17.94 \\
59  & 0 56 10.55 &  -1 04 04.20  &  0.048 &  0.283 &  22.19&  99.00&  20.36&  20.33&  19.68&  19.50&  19.48&  19.06&  18.78&  18.70&  18.91&  18.72&  18.72&  18.55&  99.00 \\
\noalign{\smallskip}\hline
\end{tabular}
\end{table}

\begin{table}
\setcounter{table}{2}
\centering

\begin{minipage}{35mm}
\caption{\it --- Continued.}\end{minipage}\vspace{0pt}

\small
\scriptsize \tabcolsep 0.70mm
\renewcommand\arraystretch{1.1}
\begin{tabular}{rccccccccccccccccccc}
\hline\noalign{\smallskip} {No.}&{R.A.} &{Decl.} & {$z_{\rm ph}$}
&{fracDeV} & {$a$} & {$b$} & {$c$} & {$d$} & {$e$} & {$f$} & {$g$} &
{$h$} & {$i$} & {$j$} & {$k$} & {$m$}
& {$n$} & {$o$} & {$p$} \\
\noalign{\smallskip}\hline\noalign{\smallskip}
60  & 0 56 59.41 &  -1 21 51.50  &  0.048 &  0.485 &  99.00&  21.11&  20.47&  20.28&  19.69&  19.51&  19.73&  18.99&  18.92&  18.86&  18.92&  18.70&  18.90&  18.40&  19.62 \\
61  & 0 54 30.41 &  -1 39 21.10  &  0.049 &  0.298 &  20.91&  19.41&  18.84&  99.00&  18.07&  17.85&  17.54&  17.42&  17.18&  17.14&  17.18&  16.93&  99.00&  16.78&  16.72 \\
62  & 0 57 12.41 &  -0 58 28.50  &  0.049 &  0.329 &  20.92&  20.73&  19.67&  19.12&  18.77&  18.76&  18.36&  18.68&  18.16&  18.16&  17.99&  17.88&  17.67&  17.81&  99.00 \\
63  & 0 54 53.93 &  -1 40 43.10  &  0.050 &  0.237 &  21.08&  21.88&  20.14&  99.00&  19.41&  19.29&  18.83&  19.10&  18.98&  18.73&  99.00&  18.70&  18.51&  18.58&  18.05 \\
64  & 0 56 06.48 &  -1 37 54.00  &  0.050 &  0.000 &  21.01&  20.27&  19.93&  99.00&  19.37&  19.39&  19.58&  19.55&  19.02&  18.94&  19.03&  18.72&  18.82&  18.35&  17.69 \\
65  & 0 56 12.34 &  -1 31 50.20  &  0.050 &  0.000 &  21.12&  20.05&  19.83&  99.00&  19.27&  19.33&  19.23&  18.99&  18.93&  18.86&  19.02&  18.75&  18.77&  18.77&  19.00 \\
66  & 0 56 13.87 &  -1 16 24.30  &  0.050 &  0.863 &  21.99&  21.02&  20.49&  21.37&  19.80&  19.65&  19.84&  19.86&  18.75&  19.18&  20.27&  18.63&  18.48&  18.79&  18.16 \\
67  & 0 55 10.34 &  -1 02 42.90  &  0.050 &  0.000 &  22.14&  21.08&  20.06&  19.73&  18.80&  18.72&  18.43&  18.61&  18.28&  18.07&  18.35&  17.90&  17.80&  17.70&  18.34 \\
68  & 0 57 55.99 &  -0 59 55.80  &  0.050 &  0.000 &  20.19&  19.39&  18.87&  18.64&  18.28&  18.20&  17.95&  17.92&  17.74&  17.59&  17.71&  17.62&  17.50&  17.48&  17.17 \\
69  & 0 54 43.20 &  -1 20 05.10  &  0.051 &  0.180 &  20.40&  19.72&  19.52&  19.08&  18.94&  99.00&  19.10&  19.38&  18.96&  18.88&  18.72&  18.67&  18.58&  18.66&  17.85 \\
70  & 0 56 07.58 &  -1 20 38.10  &  0.051 &  1.000 &  19.72&  18.90&  18.32&  17.83&  17.35&  17.16&  16.91&  16.79&  16.56&  16.55&  16.51&  16.27&  16.19&  16.06&  16.02 \\
71  & 0 55 15.35 &  -0 54 59.90  &  0.051 &  0.000 &  21.42&  19.98&  19.46&  19.11&  18.51&  18.61&  18.23&  18.42&  18.07&  18.06&  18.16&  17.75&  17.65&  17.65&  17.54 \\
72  & 0 56 35.20 &  -1 07 10.90  &  0.052 &  0.271 &  21.61&  20.42&  20.08&  19.97&  19.36&  19.21&  18.82&  18.64&  18.63&  18.38&  18.58&  18.32&  18.21&  18.34&  18.30 \\
73  & 0 54 44.56 &  -1 28 19.30  &  0.053 &  0.000 &  20.29&  20.02&  19.26&  18.68&  18.57&  18.22&  17.88&  17.83&  17.68&  17.60&  17.63&  17.48&  17.28&  17.25&  17.02 \\
74  & 0 56 34.55 &  -1 25 46.20  &  0.053 &  0.586 &  20.67&  20.06&  19.54&  19.10&  18.69&  18.60&  18.07&  17.99&  18.04&  17.84&  17.89&  17.78&  17.84&  17.55&  17.67 \\
75  & 0 56 53.33 &  -1 13 24.70  &  0.053 &  0.000 &  21.79&  20.44&  20.43&  99.00&  19.44&  19.57&  19.12&  18.92&  19.10&  18.85&  19.18&  18.88&  18.73&  18.62&  18.72 \\
76  & 0 57 13.58 &  -1 03 16.30  &  0.053 &  0.271 &  20.23&  19.39&  18.97&  18.67&  18.38&  18.37&  18.16&  18.18&  18.02&  17.98&  17.99&  17.90&  17.73&  17.82&  18.57 \\
77  & 0 55 33.50 &  -1 29 00.80  &  0.054 &  0.784 &  20.49&  19.77&  19.40&  18.91&  18.55&  18.40&  18.09&  17.97&  17.77&  17.69&  17.66&  17.42&  17.27&  17.31&  17.37 \\
78  & 0 57 05.14 &  -0 50 24.90  &  0.054 &  0.413 &  22.61&  21.81&  20.13&  19.99&  19.25&  19.23&  18.74&  18.80&  18.47&  18.50&  18.60&  99.00&  18.26&  18.39&  17.59 \\
79  & 0 56 34.49 &  -0 53 35.80  &  0.056 &  0.998 &  21.03&  20.41&  20.33&  19.73&  18.85&  18.87&  18.70&  18.41&  18.22&  18.15&  17.93&  17.79&  17.71&  17.47&  17.40 \\
80  & 0 55 40.63 &  -1 36 18.20  &  0.057 &  0.514 &  20.88&  19.97&  19.58&  99.00&  18.75&  18.55&  18.20&  18.18&  18.02&  17.98&  17.94&  17.77&  17.68&  17.62&  99.00 \\
81  & 0 55 37.21 &  -1 35 30.30  &  0.057 &  0.769 &  19.89&  19.22&  18.70&  99.00&  17.82&  17.64&  17.40&  17.22&  17.03&  16.91&  16.91&  16.72&  16.65&  16.57&  16.40 \\
82  & 0 55 31.75 &  -1 34 57.60  &  0.057 &  0.094 &  20.82&  20.22&  19.67&  99.00&  18.95&  18.97&  18.64&  18.61&  18.58&  18.38&  18.68&  18.36&  18.36&  18.44&  18.37 \\
83  & 0 56 29.31 &  -1 05 56.50  &  0.057 &  0.780 &  20.36&  19.47&  19.02&  18.45&  18.14&  18.01&  17.75&  17.57&  17.44&  17.54&  17.30&  17.19&  17.10&  17.01&  16.96 \\
84  & 0 55 05.80 &  -0 53 40.30  &  0.057 &  0.033 &  20.32&  19.71&  19.33&  18.63&  18.31&  18.13&  17.85&  17.68&  17.54&  17.47&  99.00&  17.24&  17.14&  17.02&  17.23 \\
85  & 0 55 39.07 &  -1 34 49.90  &  0.058 &  1.000 &  20.97&  20.24&  19.66&  99.00&  18.80&  18.74&  18.43&  18.42&  18.13&  17.97&  17.93&  17.74&  17.69&  17.62&  17.47 \\
86  & 0 56 59.62 &  -1 15 54.20  &  0.059 &  0.543 &  20.41&  20.24&  19.46&  18.96&  18.50&  18.46&  18.09&  17.88&  17.82&  17.90&  17.68&  17.55&  17.42&  17.43&  17.25 \\
87  & 0 57 45.14 &  -1 44 54.00  &  0.060 &  0.018 &  20.77&  20.20&  19.48&  19.02&  17.93&  18.18&  99.00&  17.38&  17.52&  18.24&  18.41&  18.19&  18.84&  17.70&  17.79 \\
88  & 0 55 40.59 &  -1 35 41.10  &  0.060 &  0.497 &  21.08&  20.17&  20.25&  99.00&  19.83&  19.38&  19.45&  19.51&  19.39&  19.39&  19.53&  19.03&  18.98&  18.73&  17.68 \\
89  & 0 57 10.97 &  -1 22 35.70  &  0.060 &  0.000 &  20.30&  19.65&  19.17&  18.87&  18.52&  18.30&  18.09&  18.05&  17.90&  17.95&  17.81&  17.57&  17.70&  17.41&  17.13 \\
90  & 0 56 08.43 &  -1 15 16.00  &  0.060 &  0.349 &  21.20&  19.83&  19.44&  18.80&  18.54&  18.36&  18.24&  18.12&  17.88&  99.00&  17.70&  17.55&  17.54&  17.38&  17.40 \\
91  & 0 57 24.61 &  -0 57 49.10  &  0.060 &  0.417 &  21.44&  99.00&  20.76&  20.34&  19.93&  19.82&  19.45&  19.31&  19.49&  19.45&  19.18&  19.05&  18.83&  19.05&  21.11 \\
92  & 0 54 50.51 &  -0 53 35.50  &  0.060 &  0.778 &  22.49&  99.00&  21.81&  21.57&  19.86&  20.15&  19.44&  19.31&  19.31&  19.40&  99.00&  18.74&  18.79&  18.55&  17.96 \\
93  & 0 55 18.68 &  -0 49 27.20  &  0.060 &  0.302 &  20.22&  19.77&  19.22&  18.72&  18.66&  18.61&  18.49&  18.44&  18.32&  18.43&  99.00&  18.00&  99.00&  17.96&  18.82 \\
94  & 0 54 39.05 &  -0 55 47.10  &  0.060 &  0.942 &  22.30&  20.81&  22.26&  23.47&  20.19&  20.68&  20.86&  19.55&  19.36&  19.90&  99.00&  19.10&  19.07&  18.75&  17.77 \\
95  & 0 56 45.13 &  -1 17 59.80  &  0.061 &  0.551 &  20.07&  19.55&  18.81&  18.37&  17.91&  17.76&  17.47&  17.32&  17.17&  17.17&  17.08&  16.90&  16.81&  16.75&  16.42 \\
96  & 0 56 49.00 &  -1 17 32.90  &  0.061 &  0.333 &  20.19&  19.55&  18.96&  18.30&  18.03&  17.81&  17.40&  17.28&  17.10&  17.03&  16.86&  16.77&  16.69&  16.59&  16.42 \\
97  & 0 54 55.40 &  -1 01 06.80  &  0.061 &  0.000 &  21.62&  20.46&  20.13&  19.91&  19.19&  19.32&  18.85&  19.37&  19.10&  19.21&  99.00&  18.63&  18.69&  18.64&  18.69 \\
98  & 0 57 00.21 &  -0 51 02.70  &  0.061 &  0.000 &  20.34&  19.60&  19.49&  19.05&  18.47&  18.44&  17.88&  18.10&  17.85&  17.86&  17.78&  17.60&  17.61&  17.41&  17.20 \\
99  & 0 57 04.09 &  -1 09 06.60  &  0.062 &  0.254 &  20.34&  19.80&  19.14&  18.78&  18.25&  18.08&  17.81&  17.68&  17.56&  17.38&  17.40&  17.24&  17.23&  17.07&  17.05 \\
100 & 0 55 10.65 &  -0 56 07.90  &  0.062 &  0.374 &  21.51&  21.35&  19.75&  19.65&  19.09&  19.05&  18.99&  18.84&  18.55&  18.75&  99.00&  18.31&  18.52&  18.21&  18.03 \\
101 & 0 54 49.28 &  -0 55 02.20  &  0.062 &  0.493 &  21.50&  21.05&  20.75&  20.31&  19.56&  19.49&  19.33&  18.66&  18.99&  18.91&  99.00&  18.44&  18.65&  18.26&  18.89 \\
102 & 0 57 28.06 &  -1 14 59.00  &  0.063 &  0.938 &  21.68&  20.72&  20.58&  20.05&  19.50&  19.40&  18.98&  18.72&  18.63&  99.00&  18.37&  18.28&  18.02&  18.18&  17.72 \\
103 & 0 55 00.15 &  -1 18 42.90  &  0.063 &  0.189 &  21.35&  22.01&  20.44&  19.81&  19.62&  99.00&  20.31&  19.75&  19.37&  19.21&  99.00&  18.81&  19.60&  18.86&  18.30 \\
104 & 0 55 36.59 &  -0 50 42.80  &  0.063 &  0.633 &  20.35&  19.13&  18.85&  18.42&  18.20&  18.07&  17.98&  17.74&  17.64&  17.57&  99.00&  17.40&  17.43&  17.26&  17.04 \\
105 & 0 54 49.48 &  -1 32 33.90  &  0.064 &  0.000 &  99.00&  21.45&  19.67&  19.62&  19.42&  19.73&  18.68&  18.98&  19.11&  20.30&  19.88&  18.91&  18.16&  20.15&  99.00 \\
106 & 0 55 57.48 &  -1 40 26.10  &  0.064 &  0.942 &  20.13&  19.37&  18.88&  99.00&  18.61&  18.55&  18.43&  99.00&  18.37&  18.53&  18.56&  18.39&  18.43&  18.58&  19.18 \\
107 & 0 57 14.62 &  -1 36 04.40  &  0.064 &  0.155 &  21.03&  21.53&  20.34&  99.00&  19.94&  20.02&  19.94&  19.01&  19.45&  19.70&  19.53&  19.44&  19.28&  19.59&  19.07 \\
108 & 0 54 38.22 &  -1 28 35.00  &  0.064 &  0.932 &  21.28&  21.34&  20.04&  19.91&  19.88&  19.52&  19.76&  19.46&  19.31&  19.20&  99.00&  19.20&  19.04&  19.27&  18.52 \\
109 & 0 55 30.97 &  -1 28 38.20  &  0.065 &  0.000 &  21.20&  20.40&  19.44&  19.18&  18.91&  18.83&  18.31&  18.50&  18.35&  18.30&  18.23&  18.29&  18.23&  18.19&  18.81 \\
110 & 0 56 28.27 &  -0 56 41.20  &  0.065 &  1.000 &  19.47&  18.76&  18.35&  17.95&  17.80&  17.72&  17.47&  17.25&  17.16&  17.17&  17.12&  16.96&  16.93&  16.91&  16.79 \\
111 & 0 56 00.93 &  -0 56 26.80  &  0.065 &  0.072 &  21.54&  20.16&  20.02&  19.50&  19.40&  19.13&  18.57&  18.73&  18.60&  18.57&  99.00&  18.43&  18.35&  18.34&  18.34 \\
112 & 0 56 40.83 &  -0 55 20.40  &  0.065 &  0.378 &  21.59&  20.64&  20.49&  20.11&  20.00&  19.74&  19.58&  19.17&  19.22&  18.95&  19.51&  18.60&  19.25&  18.71&  17.61 \\
113 & 0 55 49.02 &  -0 52 37.10  &  0.065 &  0.632 &  21.41&  21.02&  20.27&  19.85&  19.41&  19.48&  19.33&  19.34&  19.03&  18.92&  19.12&  18.86&  18.69&  18.88&  18.15 \\
114 & 0 58 02.90 &  -1 26 15.70  &  0.066 &  0.023 &  20.27&  19.58&  19.37&  19.13&  18.85&  18.68&  18.45&  18.35&  18.21&  18.15&  18.19&  99.00&  18.01&  17.91&  17.66 \\
115 & 0 55 55.93 &  -1 17 50.20  &  0.066 &  0.956 &  20.51&  19.79&  19.25&  18.72&  18.39&  18.22&  17.83&  17.72&  17.57&  17.62&  17.54&  17.36&  17.38&  17.18&  17.06 \\
116 & 0 54 50.14 &  -0 54 58.70  &  0.066 &  0.485 &  20.94&  21.06&  20.47&  19.86&  19.17&  18.93&  18.71&  18.41&  18.22&  18.07&  99.00&  17.72&  17.76&  17.61&  17.16 \\
117 & 0 57 10.32 &  -0 48 57.60  &  0.066 &  0.352 &  22.41&  20.33&  20.05&  19.71&  19.51&  19.08&  18.60&  18.61&  18.61&  18.39&  18.77&  99.00&  99.00&  18.28&  20.28 \\
\noalign{\smallskip}\hline
\end{tabular}
\end{table}

\section{The Properties of the Cluster A119}
\label{sect:analysis}
\subsection{Spatial Distribution and Localized Velocity Structure}
Based on sample II, we investigate spatial distribution and localized
velocity structure of A119. Left panel of Fig.~8 shows contour maps
of surface density with smoothing Gaussian window of
$\sigma=2\arcmin$, superimposed projected distribution of galaxies in
sample II. The early-types and late-types are marked with open
circles and asterisks, respectively. Density contour for sample II
appears to have more significant deviation from spherically symmetry,
indicating that A119 is a dynamically complex system with significant
substructures.

Three substructures (A, B, and C) shown in Fig.~3 are confirmed in
Fig.~8. Result of the $\kappa$-test for sample II is given in
Table~4. The probability P($\kappa_n>\kappa_n^{\rm obs}$) is less
than $1\%$, strongly suggesting a more significant detection of
substructure. The bubble clustering in clump A appears to be split
into two bunches, which makes the picture of merging along the line
of sight clearer. In addition, two bunches of bubbles are newly
detected at about 23\arcmin southwest and northwest of the main
concentration. Considering the large uncertainties of $z_{ph}$
estimate, follow-up spectroscopy is needed to confirm these two
substructures.

\begin{figure}[!h]
\centering
\includegraphics[width=70mm,height=70mm,angle=-90]{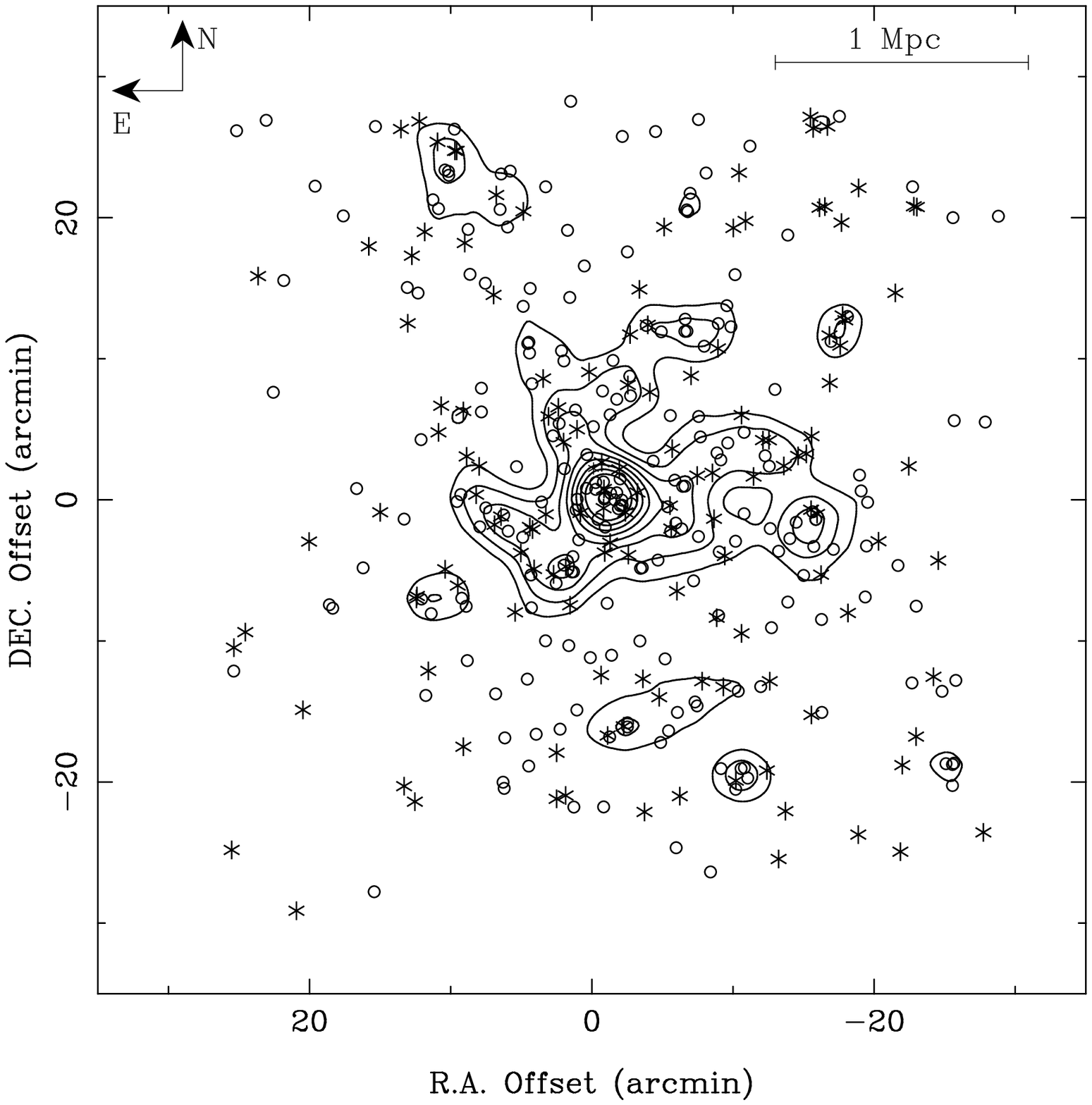}
\includegraphics[width=70mm,height=70mm,angle=-90]{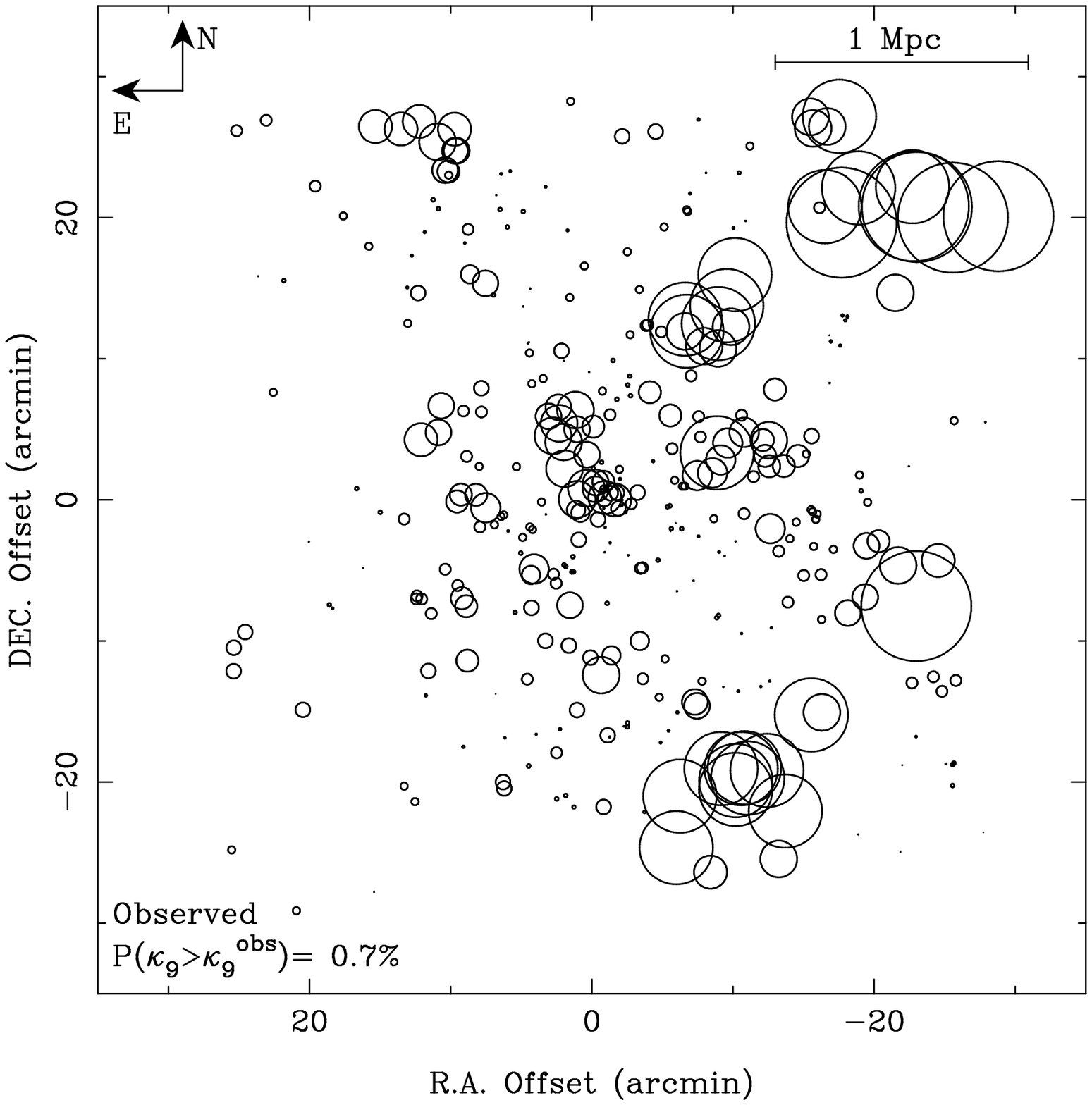}
\caption{ \baselineskip 3.6mm left: Spatial distribution for 355
member galaxies in sample II. The smoothing Gaussian window is
$\sigma$=2\arcmin, the contour levels are 0.22, 0.32, 0.42, 0.52,
0.62, 0.72, 0.82, and 0.92 arcmin$^{-2}$. open circles for early-type
galaxies and asterisks for late-type galaxies, respectively.
 Right: Bubble plot showing the localized
variation for groups of the 9 nearest neighbors in sample II.
\label{fig8}}
\end{figure}

\begin{table}[h!!]
\vspace{0pt} \centering
\begin{minipage}{115mm}
\caption[]{$\kappa$-test for member galaxies in samples I and II of
A119}
\end{minipage}

\begin{tabular}{crc}
\hline\noalign{\smallskip}
{Neighbor size} & Sample I & Sample II \\
$n$ & $P$($\kappa_n>\kappa_n^{obs}$) & $P$($\kappa_n>\kappa_n^{obs}$)\\
\noalign{\smallskip}   \hline \noalign{\smallskip}
 6& 14.1\% & $<$0.1\% \\
 7& 14.6\% & $<$0.1\% \\
 8& 13.9\% &    0.2\% \\
 9&  7.7\% &    0.7\% \\
10&  8.5\% &    1.1\% \\
11&  9.4\% &    0.6\% \\
12& 11.1\% &    0.5\% \\
\noalign{\smallskip}   \hline\noalign{\smallskip}
\end{tabular}
\end{table}

\subsection{Morphology and Luminosity Segregations}

The clustercentric distance $R$ and local galaxy density are traditional
parameters tracing the environment in a cluster. Projected local galaxy
density is commonly defined as $\Sigma_{10} = 10/({\pi r^2_{10}}) $, where
$r_{10}$ is the distance from a given galaxy to the ninth nearest neighbor
(\citealt{dressler80}). As reviewed by \citet{sandage05}, morphological
classification is the most intuitive tool for extragalactic astronomy.
\citet{dressler80} investigated the relation between morphology and
local galaxy density, so called morphology-density relation, and
found that fraction of spiral galaxies decrease with increasing local
density for low-redshift galaxy clusters. This relation has been
studied by other authors subsequently (\citealt{hash99,goto03}).
\citet{whitmore91,whitmore93} re-analyzed the samples of
\citet{dressler80}, and argued that the correlation between
morphology and clustercentric radius $R$ appears tighter than
morphology-density relation. \citet{sanroma90} and \citet{whitmore93}
suggested that the global parameter, clustercentric radius $R$,
should be more fundamental. Local and global processes in cluster
galaxies are generally considered to be two causes of different
morphologies. Due to the close relation between local galaxy density
and clustercentric radius(\citealt{beers86,merrifield89}), the
argument on which parameter is more fundamental in morphology and
luminosity segregation is still inconclusive so far.

For checking the presence of luminosity segregation in A119,
\citet{pracy05} studied the luminosity functions and locations
of cluster galaxies in A119 on the basis of their V-band photometry.
The information of radial velocities were not taken into account during
their sampling. They found that the core radius of a King profile is
invariant with intrinsic luminosity. The luminosity functions for member
galaxies within three annuli ($r<0.3$Mpc, $0.3<r<0.6$Mpc, and
$0.6<r<1.5$Mpc) are fitted with the Schechter function, and no
significant systematic correlation with cluster-centric radius are found.
Alternatively, \citet{driver98} defined the dwarf-to-giant ratio (DGR)
to quantify the luminosity distribution, and found that giant galaxies
are more centrally concentrated than the dwarfs in galaxy clusters. In
our paper, because of the limited number of member galaxies, we define
the faint-to-bright ratio (FBR) to describe the luminosity distribution:
FBR $=$ ${\sum N(M_R>-19.5)}/{\sum N(M_{R}<-19.5)}$,
where $M_R$ is the absolute magnitude for the conventional
Kron-Cousins $R$ band which can be calculated via the equations in
\citet{zhou03a}.

\begin{figure}[th]
\centering
\includegraphics[width=70mm,height=70mm]{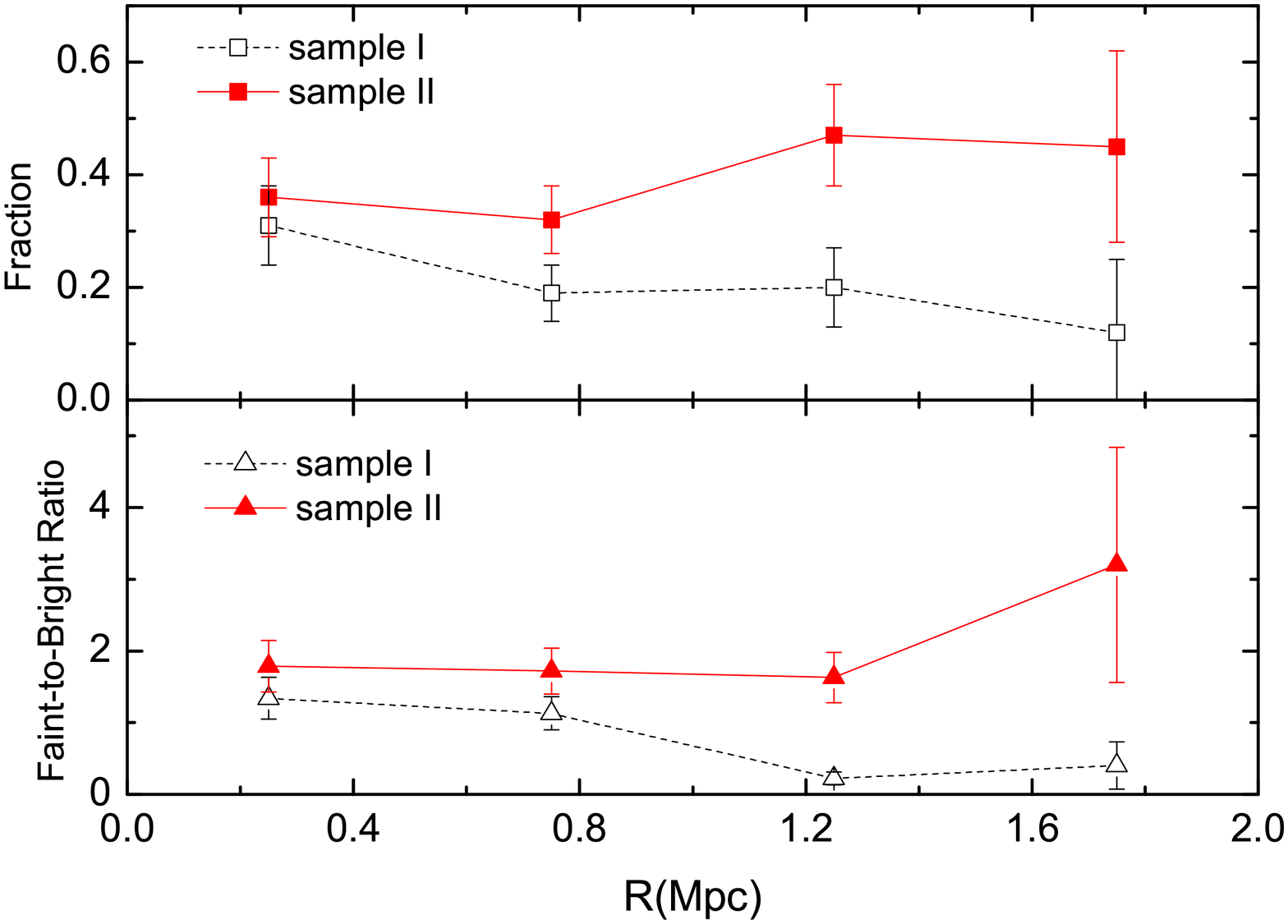}
\includegraphics[width=70mm,height=70mm]{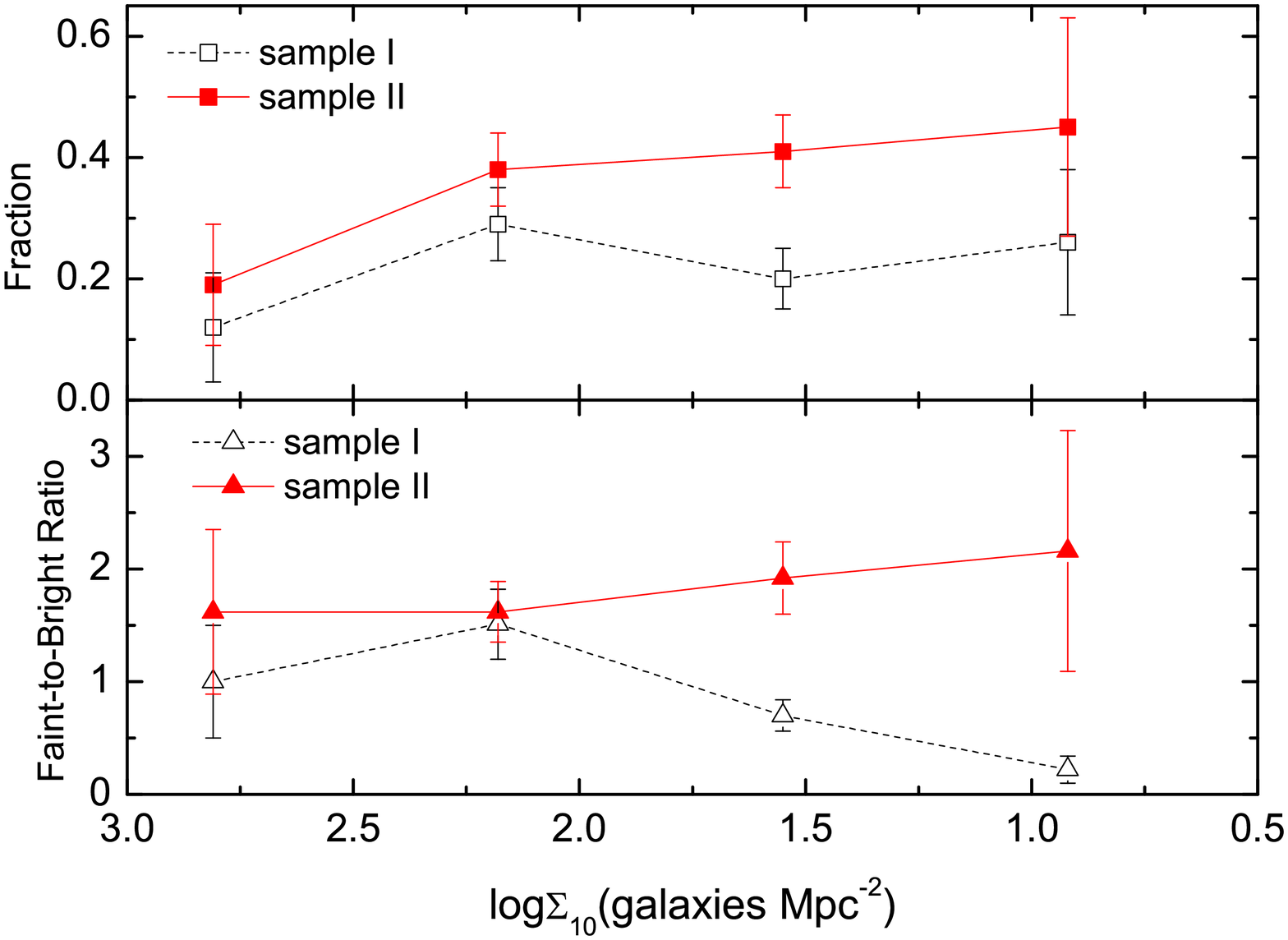}
\caption{ \baselineskip 3.6mm Left: Fraction of late-type
galaxies and faint-to-bright ratio (FBR) as functions of clustercentric
radius for sample I and sample II, respectively. Right: Fraction of
late-type galaxies and faint-to-bright ratio (FBR) as functions of
local galaxy density for sample I and sample II, respectively. The width
of annuli is 0.5 Mpc, and bin size of local density is
$\Delta$log$\Sigma_{10} = 0.63$ (galaxies Mpc$^{-2}$).}
\label{fig9}
\end{figure}

Fig.~9 shows the fraction of late-type galaxies and the FBR as functions
of clustercentric radius $R$ and local density $\Sigma_{10}$ for both
samples. For sample I, the late-type galaxy fraction slightly decreases
with increasing $R$,
which might be due to the strong bias in sample I. As a matter of fact,
the spectroscopic redshifts in sample I are contributed by several
observations (see Table 2), and the bright early-type galaxies closer to
the cluster center have greater probability to be selected as
spectroscopic targets. Small number of faint member galaxies are
spectroscopically detected in the outskirt regions.

Some faint member galaxies in the outskirt low-density region are selected by
our multicolor photometry, which results in a prominent increase of the
late-type galaxy fraction and the FBR for sample II in the regions
with larger $R$ and lower density (see the red points in Fig.~9).
The FBR and late-type fraction in sample II appears to increase
monotonically with decreasing local density, rather than with
increasing clustercentric radius. However, owing to the limited number
of member galaxies, significant uncertainties in the late-type
galaxy fraction and the FBR can be found for the low-density and outskirt
regions in sample II, which will surely reduce the probability of
monotonically increasing tendency. Considering the error bars in the right
panels of Fig.~9,
we apply Monte Carlo simulations to estimate the probabilities
of monotonic increase for sample II. As a result, the late-type galaxy
fraction has a probability of 31.0\% to monotonically increase with the
decreasing density, and the probability that the FBR monotonically increases
with decreasing density is about 18.8\%. Therefore, we may safely conclude
that no clear evidences are found for morphology and luminosity segregations
in A119. A deep and complete sample of member galaxies is necessary to
further determine which environmental indicator is more fundamental in
the future study.

\section{Star Formation Properties of Cluster Galaxies }
The star formation properties of member galaxies in a cluster can
help us to understand formation and evolution of galaxies and their
host cluster. Thus, This is important to observe the systematic
tendency of the star formation properties for the galaxies in a
cluster.

With the evolution synthesis model, PEGASE (version 2.0,
\citet{fioc97,fioc99}), we investigate the star formation properties
of A119. A Salpeter initial mass function (IMF)
(\citealt{salpeter55}) is adopted. The star formation rate (SFR) are
assumed to be in exponentially decreasing form, $SFR(t) \propto
e^{-t/{\tau}}$, where the time scale $\tau$ ranges from 0.5 to 30.0
Gyr. To avoid the degeneracy between age and metallicity in the
model, the same age of 12.7 Gyr is adopted for all the member
galaxies in A119, responding to the age of the first generation stars
at $z_{\rm c}$= 0.0442. A zero initial metallicity of interstellar
medium (ISM) is assumed. Firstly, a series of modelled spectra at
rest frame ({\it z}=0) with various star formation histories are
generated by running the PEGASE code. Then we shift them to the
observer's frame for a given redshift, and convolve with the
transmission functions of the BATC filters. As a result, we obtain
the template SED library for the BATC photometric system. Based on
the template SED library, the best fit (in the $\chi^{2}$ sense) to
the observed SEDs are performed for 238 member galaxies with known
spectrocopic redshifts. The SFR time scale ($\tau$), mean ISM
metallicity (Z$_{\rm ISM}$), and the mean stellar age ($t_{\star}$)
weighted by mass and light for each cluster galaxy can be achieved.

\begin{figure}[!b]
\centering
\includegraphics[width=140mm,height=140mm,angle=-90]{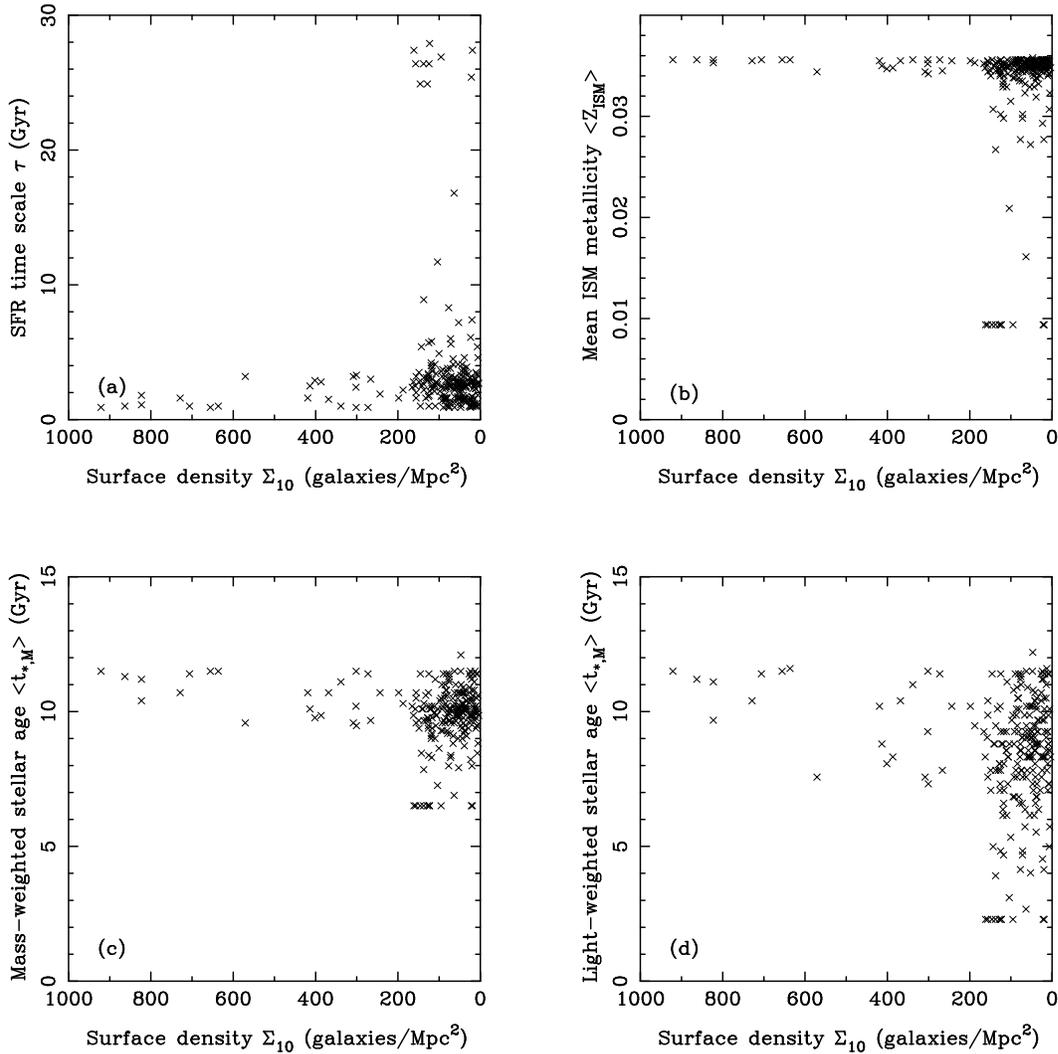}
\caption{Star formation properties, such as the SFR time scale
$\tau$, mean ISM metallicity, and the mean stellar ages weighted by
mass and light, for the galaxies with known {\it z$_{sp}$} values as
the functions of local galaxy density. \label{fig10}}
\end{figure}

As we know, early-type galaxies in the field commonly have low star
formation rates, whereas late-type galaxies have high star formation
activity. This bimodality is modified in dense environments: a higher
occurrence of passive spiral galaxies is found in clusters than in
the field, whereas star-forming elliptical galaxies are rarer in
clusters (Bamford et al. 2009). This suggests that star formation
rate couples most strongly to environment, with morphology being only
a secondary correlation. Firstly, we attempt to find the tendency of
star formation property along the clustercentric radius $R$ of A119.
However, no any tendencies are found as expected. This suggests
clustercentric radius is not a good environmental indicator for A119.
An alternative explanation is that local processes (e.g.,
galaxy-scale interaction) affect star formation activities, rather
than the global processes (e.g., ram-pressure stripping within
cluster-scale environment).
This result proves the correctness of the morphology-density relation
pointed out by \citet{dressler80} again, which can be well explained
in the context of hierarchical cosmological scenario
(\citealt{poggianti04}).

Fig.~10 shows the star formation properties as a function of the
local galaxy density for 238 member galaxies in sample I.
We take $\Sigma_{10} = 200\,{\rm Mpc}^{-2}$ as a boundary between the
regions with high and low densities. Panel (a) shows that the
galaxies in the high-density regions have shorter SFR time scale than
those in the low-density regions. Panel (b) shows that the galaxies
in high-density regions are more likely to have a higher metallicity
of interstellar medium (ISM). It is generally regarded that the
galaxies in high-density regions (e.g., cluster core) tend to be more
luminous and massive. This trend can be commonly interpreted as
luminosity-metallicity relation and mass-metallicity
relation(\citealt{garnett87}). Panels (c) and (d) show that the
galaxies in low-density regions tend to possess younger stellar
population with shorter mean stellar ages weighted by either mass or
light. Variation of light-weighted stellar age is remarkably
widespread, particularly in the high-density regions, in contrast to
mass-weighted stellar age. A possible explanation is that current SFR
in a cluster is mainly contributed by the late-type galaxies in low
density regions, and young stellar population has a greater weight in
average age calculation. It is reasonable that the light-weighted
mean of stellar ages tend to be younger for cluster galaxies.

\begin{figure}[t]
\centering
\includegraphics[width=70mm,height=140mm,angle=-90]{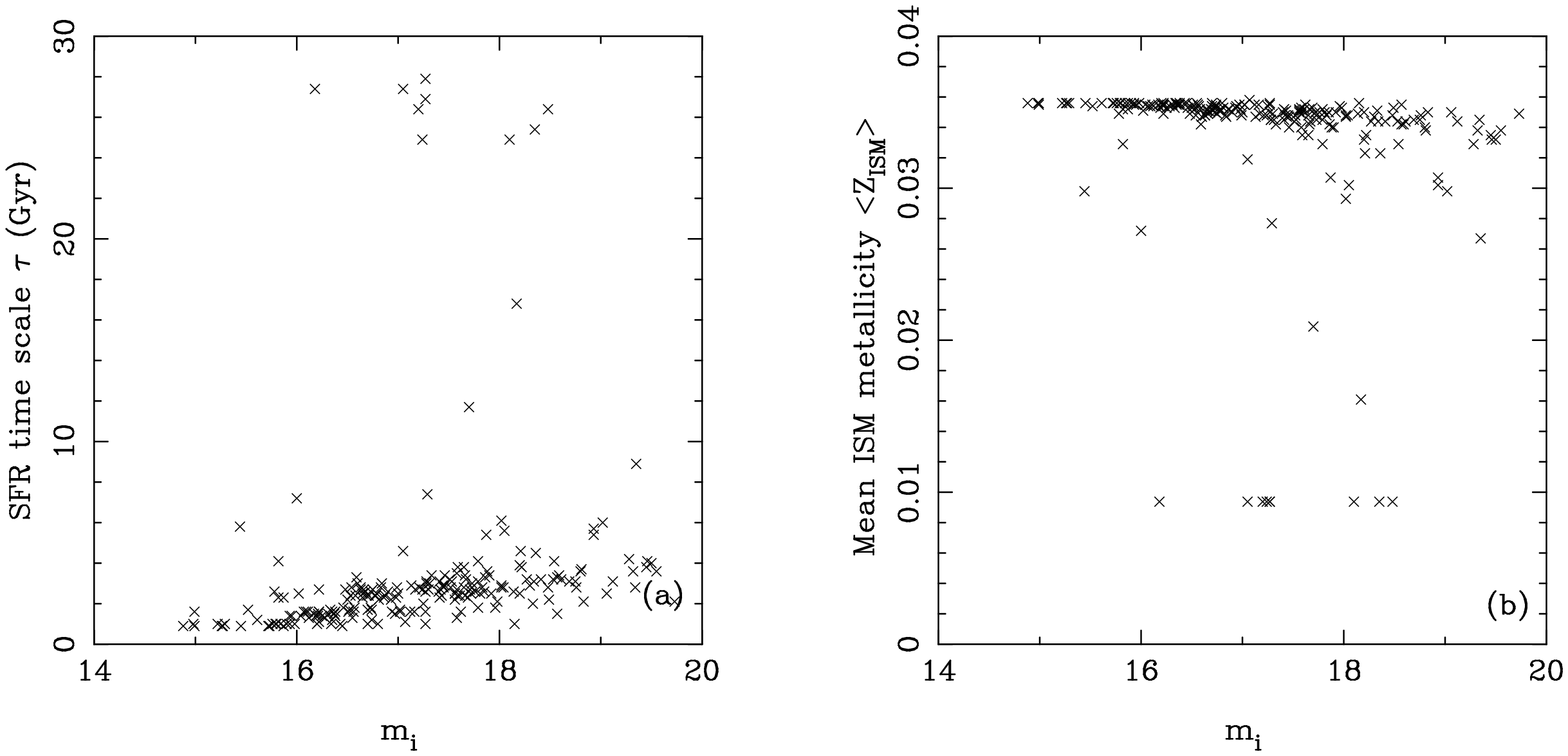}
\caption{The SFR time scale $\tau$ and mean ISM metallicities for the
galaxies with known {\it z$_{sp}$} in A119 against magnitude in the
BATC $i$ band. \label{fig11}}
\end{figure}

It is generally considered that member galaxies in a cluster have
same distance modulus, apparent magnitude could reflect their
intrinsic luminosity. The SFR time scale ($\tau$) and mean ISM
metallicity for the galaxies in sample I are illustrated in Fig.~11,
as the functions of apparent magnitude in the BATC $i$ band.
Fig.~11(a) shows brighter galaxies tend to have shorter SFR time
scales. Massive and luminous member galaxies located at high-density
region began to fall into gravitational potential well at earlier
time, thus their star formation activities have been reduced by some
physical processes for a longer time, which results in a short
timescale of star formation. Fig.~11(b) gives the variation of mean
ISM metallicities with their magnitudes in the BATC $i$ band. Fainter
member galaxies tend to have lower mean ISM metallicities, while
luminous galaxies are likely to have greater metallicities. This is
consistent with the ideas that metals are selectively lost from faint
galaxies with shallow potential wells via galactic winds
(\citealt{melbourne02,tremonti04}).

\section{Summary}
\label{sect:summary}

X-ray observations suggest that the nearby cluster A119 is not a
regular and well-relaxed cluster. We present our multicolor
photometry in optical bands for this galaxy cluster, on the basis of
the Beijing-Arizona-Taiwan-Connecticut(BATC) 15 intermediate filters
system that covers almost whole optical wavelength domain. We obtain
the SEDs of 1376 galaxies brighter than $i_{BATC}=19^m.5$ in our
viewing field of $58\arcmin \times58\arcmin$. There are 368 galaxies
with available spectroscopic redshifts, among which 238 galaxies with
10736 \kms $< cz_{sp} <$ 15860 \kms are regarded as member galaxies
of A119 (sample I).

Based on sample I, both projected distribution and localized velocity
structure support the picture that A119 is a dynamically young
cluster with some substructures, which is in agreement with the X-ray
image. Three potential substructures are confirmed in localized
deviation of velocity distribution in the central region of A119.
Clump A is found to have a very large velocity dispersion, which
supports a merger at the cluster center along the line of sight.
Clump C might be a compact group of galaxies which have a bulk
velocity of 12966 \kms.

Photometric redshift technique is applied to the faint galaxies
without $z_{sp}$ values, and the CM relation for early-type galaxies
is also used to constrain the membership selection. As a result, 117
faint galaxies are selected as candidates of member galaxies. An
enlarged sample of 355 member galaxies, called sample II, is obtained
by combining with sample I. Based on sample II, projected
distribution and localized velocity structure are investigated. The
result of $\kappa$-test for sample II definitely suggests significant
substructures in A119. Three substructures mentioned above are all
enhanced in bubble plot.

Subsequently, morphology and luminosity segregations on the basis of
sample II are investigated. We define the faint-to-bright ratio to
quantify the luminosity distribution, and find that the fraction
of late-type galaxy and the faint-to-bright ratio have very
small probabilities to monotonically increase with decreasing local
galaxies density. No significant evidences for morphology and
luminosity segregations are found.

With an evolutionary synthesis model, PEGASE, star formation
properties of sample I is studied. Environmental effect on star
formation histories is found for these member galaxies. The bright
massive galaxies in the high-density region of A119 are found to be
more likely to have shorter SFR time scales, higher mean ISM
metallicities and longer mean stellar ages, and vice versa. These
results can be well interpreted by the existing correlations, such as
the morphology-density relation, the luminosity-metallicity relation,
and the mass-metallicity relation.

\normalem
\begin{acknowledgements}
We thank the anonymous referee for his/her invaluable comments and suggestions.
This work was funded by the National Natural Science Foundation of
China (NSFC) (Grant Nos.~11173016, 10873016, 11073032, 11003021, and 10803007),
and by the National Basic Research Program of China (973 Program)
(Grant No.~2007CB815403). We would like to thank Prof. Kong, X. and
Cheng, F.-Z. at the University of Science and Technology of China for
the valuable discussion.

This research has made use of the NED, which is operated by the Jet
Propulsion Laboratory, California Institute of Technology, under
contract with the National Aeronautics and Space Administration.
Funding for the SDSS was provided by the Alfred P. Sloan Foundation,
the Participating Institutions, the National Science Foundation, the
U.S. Department of Energy, the National Aeronautics and Space
Administration, the Japanese Monbukagakusho, the Max Planck Society.
The SDSS web site is http://www.sdss.org.

\end{acknowledgements}

\label{lastpage}

\end{document}